\documentclass[usenatbib]{mnras}

\usepackage[T1]{fontenc}
\usepackage{ae,aecompl}

\usepackage{graphicx}	
\usepackage{amsmath}	
\usepackage{amssymb}	

\title[Kozai--Lidov Disc Instability]{Kozai--Lidov Disc Instability}

\author[Stephen H. Lubow and Gordon I. Ogilvie]
{Stephen H. Lubow$^{1}$ \thanks{E-mail: lubow@stsci.edu}
and Gordon I. Ogilvie$^{2}$
\\
$^1$Space Telescope Science Institute, 3700 San Martin Drive, Baltimore, MD 21218, USA \\
$^{2}$ Department of Applied Mathematics and Theoretical Physics, University of Cambridge, \\Centre for Mathematical Sciences, Wilberforce Road, Cambridge CB3 0WA, UK
}

\date{Accepted April 20, 2017. Received April 18, 2017; in original form February 27, 2017}

\pubyear{2015}

\begin{document}
\label{firstpage}
\pagerange{\pageref{firstpage}--\pageref{lastpage}}
\maketitle

\begin{abstract}
Recent results by \cite{Martin14} showed  in 3D SPH simulations that tilted discs in binary
systems can be unstable to the development of global, damped Kozai--Lidov (KL) oscillations in which  the discs exchange tilt for eccentricity.
We investigate the linear stability of KL modes for tilted inviscid discs under the approximations
that the disc eccentricity is small and the disc remains flat. 
By using 1D equations, we are able to probe regimes
of large ratios of outer to inner disc edge radii that are realistic for binary systems of hundreds of AU 
separations and are not easily probed by multi-dimensional simulations. 
For order unity binary mass ratios, KL instability is possible for a window of disc aspect ratios  $H/r$ in the outer parts of a disc that roughly scale as
 $(n_{\rm b}/n)^2 \la H/r \la n_{\rm b}/n$, for binary orbital frequency $n_{\rm b}$ and  orbital frequency $n$
 at the disc outer edge.  
 We present a framework for understanding the zones of instability based on the determination
 of branches of marginally unstable modes.
   In general, multiple growing eccentric KL modes can be present in a disc.  Coplanar apsidal-nodal
precession resonances delineate instability branches. 
 We determine the range of tilt angles for unstable modes as a function of disc aspect ratio.
 Unlike the KL instability
for free particles that involves a critical (minimum) tilt angle, disc instability is possible for any nonzero tilt angle  depending on the
disc aspect ratio.
\end{abstract}




\section{Introduction}

Discs in binary systems are sometimes misaligned with respect
to their binary orbital planes. Misalignment is expected to be more likely in wider binaries with separations greater than $\sim 100$ AU
where the effects of tidal dissipation
in the disc
 are weaker and act on longer timescales \citep{Papaloizou95, Bate00, Lubow00, King13}.
There are several ideas on how noncoplanarity could come about in a young binary star system.
Noncoplanarity could be the result of initial conditions.
For example, if a young binary star system accretes turbulent gas, a second generation of accreted gas is likely to be misaligned with the binary orbit and result in misaligned discs around the young stars \citep{Bate10}.
Alternatively,
a coplanar disc
may evolve to a noncoplanar state due to an instability, such as
radiation warping \citep{Pringle96,  Ogilvie01a}.

Observational evidence for disc misalignment has been found in several binaries.
 In binary stars separated by greater than 40 AU,
misaligned discs may occur because the stellar equatorial inclinations, based on spins, are observationally inferred to be misaligned with respect to the binary orbital planes \citep{Hale94}.  More direct evidence comes from images of discs.
The young binary system HK Tau  with binary separation $\sim 400$ AU provides
direct evidence for noncoplanarity, since discs are observed around
both components with one disc edge-on and the other more face-on
\citep{Stapelfeldt98}. Recent ALMA observations by \cite{Jensen14}
suggest that its discs are mutually misaligned by more than $60^{\circ}$, although
the plane of the binary orbit is not known. Strong mutual disc misalignment ($\sim 72^{\circ}$) has lately also been detected for the two circumstellar discs in V2434 Ori, a binary system in Orion  with a similar binary separation \citep{Williams14}.

Test particles that reside  in orbits that are inclined to the orbital plane of a circular binary star
 can undergo the
powerful effects of Kozai--Lidov (KL) oscillations \citep{Kozai62, Lidov62}.  These oscillations cause particles to exchange
inclination for eccentricity. The particle initially
gains eccentricity while reducing its orbital tilt and later loses eccentricity while gaining orbital tilt back to its
original value. The process repeats in a periodic manner. For KL oscillations to occur, the particle orbit must be misaligned by more than about $39.2^{\circ}$
and less than $141.8^{\circ}$ with respect to the binary orbital plane.
 The amount of eccentricity gain can be quite large. For example, an initially circular  orbit at an inclination of $60^{\circ}$ achieves a maximum eccentricity of about 0.75 during a KL cycle. 
Further extensions of this theory show that even stronger effects can occur for eccentric orbit binaries \citep{Ford00, Lithwick11, Naoz13, Teyssandier13, Li14, Liu15}.
The KL effect for ballistic objects has been applied to a wide range of astronomical processes. These range from inclinations of asteroids and irregular satellites \citep{Kozai62, Nesvorny03} to tidal disruption events \citep{Chen11}, and the formation of Type Ia supernovae \citep{Kushnir13}.

In a recent paper, \cite{Martin14} found in SPH simulations that a fluid disc can undergo
large-scale, coherent, damped KL oscillations. In a subsequent paper \cite{Fu15a} found that these
oscillations can occur over a broad range of disc and binary parameters.
Disc self-gravity can suppress KL oscillations \citep{Fu15b}, although the suppression typically requires
that the disc conditions are not far from gravitational instability.
Such oscillations would have important consequences on the disc tilt and eccentricity evolution.
They may even result in the formation of planets,  as a consequence of the strong compression
of the disc gas due to the strong shocks produced in a highly eccentric disc \citep{Fu16}.

The geometry of a KL disc is somewhat complicated. The disc undergoes
nodal and apsidal precession, along with tilt and eccentricity oscillations.
Such a configuration is challenging to simulate with Eulerian grid-based codes because adequate resolution
would require that there be many grid cells, most of which are empty at any instant in time. KL discs are more
easily studied with a Lagrangian code, such as SPH.
However, the SPH simulations have certain limitations.
First they are subject to effects of artificial viscosity. These effects can be partially
controlled by including many particles, typically more than $1 \times 10^6$. However, simulations with such a large number of particles
require long running times. 
During the course of the simulations, viscosity causes the shape of the density
distribution to evolve and the particle count to decrease, thereby reducing the resolution.
Ideally, one would like to understand how the evolution
would occur under conditions of low viscosity, in order to isolate the effects of viscosity
from pressure. 
Another issue is the limited dynamic range of the discs studied in the SPH simulations.
Typically, the ratio of the outer to inner disc radii is about a factor of 10. For a disc in a 
several hundred AU binary, this ratio is  orders of magnitude larger.

One approach to overcoming the limitations of multidimensional simulations is to apply 1D  reduced equations based
on asymptotic methods that assume that the disc is thin: $H/r \ll 1$. Such equations have been developed in \cite{Ogilvie01} that describes the nonlinear evolution of eccentric discs and \cite{Ogilvie99, Ogilvie06} that describe the nonlinear evolution of warped discs.
Unfortunately,  there are not currently corresponding equations that describe the nonlinear evolution of both warped and eccentric discs as can occur in the case of KL discs. 

Instead, we analyze in this paper the onset of the KL oscillations  when the eccentricity is small and nonlinear effects can be ignored. 
The early states of a KL particle oscillation can be understood as an exponential growth of eccentricity resulting from a linear instability of a tilted circular orbit \citep{Tremaine14}.  It is the generalization of this instability to a continuous disc that we describe in this paper.
The analysis is carried out as a linear stability problem of an inviscid disc.
We are interested in determining the conditions under which KL oscillations can occur.
The initial stages of KL oscillations are dominated by eccentricity changes over very small changes in disc tilt. Consequently, only the eccentricity evolution needs to be followed.
In addition, we assume the disc remains flat (unwarped).
The discs found in the SPH simulations are typically quite flat, although some warping by amounts of
order the disc aspect ratio are found during the course of KL oscillations \cite[see Fig. 5 of][]{Fu15a}.
The flatness can be maintained by bending waves communicated by pressure in the disc.
The condition for flatness is that the sound crossing time is shorter than
the warping timescale, which is of order the precession timescale at the outer edge of the disc
\citep{Papaloizou95, Larwood97, Lubow00} . 

To carry out the analysis, we apply a linear 1D eccentricity evolution equation of \cite{Teyssandier16} and augment
that equation with terms that describe the effects of KL oscillations.
In a recent preprint, \cite{Zanazzi16} took a very similar approach. Our results are in agreement with
theirs where they overlap. Our analysis is different in that
we map out the regimes of instability and analyze the consequences
of having a small inner disc edge radius.

The outline of the paper is as follows.  In Section 2, we derive the equations
for the eccentricity evolution of a KL disc.  Section 3 describes the parameters
of the disc models that we analyze. Section 4 describes  some results  for the time evolution
of eccentricity and the methods used in this paper. Section 5 discusses the role of resonances in  determining KL instability. Section 6  describes the states of marginal stability for the disc models. Section 7 discusses how these states of marginal stability
are related to zones of instability. Section 8 describes  some properties of the eccentric modes, including the behaviour at small radii,  the sensitivity of the results to the inner  boundary location, and  an  estimate for the scaling of  the level of nonlinearity with eccentricity. Section 9 contains a summary.

\section{Derivation of Linearized KL Equation}

\subsection{Particle Evolution Equation for Small Eccentricity \label{sec:partev}}
We derive the linear evolution equations for a nearly circular fluid disc by first considering
the standard secular KL particle evolution equations in the quadrupole approximation \cite[e.g.,][]{Kiseleva98}.
Consider a circular orbit binary with orbital radius $a_{\rm b}$. The particle orbits about a binary member with mass $M_1$ and is gravitationally perturbed  by a companion member  with mass $M_2$ on a circular orbit. To lowest order in particle eccentricity $e$,
the secular evolution equations are given by
\begin{eqnarray}
\frac{d a}{dt} &= &0, \\
\frac{d e}{dt} &= & \frac{15 G M_2}{8 n a_{\rm b}^3}  e \sin^2{i} \sin{2 \omega}, \label{dedt}\\
\frac{d i}{dt}  &= & 0,\\
\frac{d \omega}{dt}  &= & \frac{3 G M_2}{8 n a_{\rm b}^3} (2 -3 \sin^2{i} +5 \sin^2{i} \cos{2 \omega}) \nonumber\\
&&- \frac{d \Omega}{dt} \cos{i},\label{domegadt}  \\ 
\frac{d \Omega}{dt}  &= & -\frac{3 G M_2}{4 n a_{\rm b}^3} \cos{i},
\end{eqnarray}
where $a, e, i,  \omega, \Omega$ and $n$ describe the semi-major axis, eccentricity, inclination, argument of periapsis, 
longitude of ascending node, and orbital frequency of a particle, respectively.
In obtaining these evolution equations, we dropped terms of order $e^2$ or higher on the RHSs.

We apply a complex eccentricity defined such that $E = e \exp{({\rm i} \, \omega)}$.
Consider a Cartesian coordinate system $(x, y)$ that lies in the plane of the orbit
such that the positive $x-$axis lies along the instantaneous line of ascending nodes. 
Complex eccentricity $E$ is related to the eccentricity vector $\bmath{e} $ that
points from the origin to periastron by 
$E= e_x + {\rm i} \, e_y$,  so that  $e_x(r,t)= Re(E(r,t))$ and $e_y(r,t)=Im(E(r,t)).$
In terms of a complex eccentricity,
Equations (\ref{dedt}) and (\ref{domegadt}) can be expressed as
\begin{equation}
\frac{d E}{dt} = \frac{3 G M_2}{8 n a_{\rm b}^3} ((2-3 \sin^2{i}) {\rm i} E + (5  \sin^2{i})  \, {\rm i}  E^*)-
\left(\frac{d \Omega}{dt} \cos{i} \right) {\rm i}  E \label{E1}.
\end{equation}

Note that the linear system $dE/dt =  {\rm i}  (A E + B E^*)$ 
with real coefficients $A$ and $B$ implies that $d^2E/dt^2 = (B^2-A^2)E$.  Complex eccentricity $E$
is unstable if $B^2>A^2$.
An initially nearly circular orbit  has 
\begin{eqnarray}
A &=& \frac{3 G M_2}{8 n a_{\rm b}^3} (2-3 \sin^2{i} + 2 \cos^2{i}) ,\\
B &=&  \frac{15 G M_2}{8 n a_{\rm b}^3} \sin^2{i},
\end{eqnarray}
implying that an instability occurs with growth rate
\begin{equation}
\lambda = \frac{3 \sqrt{2}G M_2}{4 n a_{\rm b}^3} \sqrt{5-3 \cos^2{i}}
\end{equation}
when $\cos^2{i} < 3/5$, in agreement with \cite{Tremaine14}.

\subsection{Disc Evolution Equation for Small Eccentricity}
We apply the above equations to a system of particles, or to a continuous eccentric disc, 
in which additional, internal forces of constraint in the normal direction maintain a rigid tilt. 
The particles then represent disc fluid elements. 
By the assumption that the disc remains flat, fluid elements then share a common $i$ and $\Omega$, 
although they have independently varying $a$, $e$, and $\omega$. 
By Newton's Third Law, the sum of the internal forces over all particles (or rings) vanishes and we have
\begin{eqnarray}
\frac{di}{dt} \int h \, dm &=& 0, \\
\frac{d \Omega}{dt} \int h \, dm &=& -\frac{3 G M_2}{4 a_{\rm b}^3} \cos{i}  \int a^2 \, dm \label{dOmdt0},
\end{eqnarray}
where $h=n a^2$  is the specific angular momentum to lowest order in $e \ll 1$ and $dm$ is a mass element of the disc.
The integrals are taken over the entire disc.
We can regard the disc inclination as constant in space and time during this small-$e$ phase.

Equation (A22) of \cite{Teyssandier16} 
provides a 1D linear secular eccentricity evolution equation of a 3D locally isothermal fluid disc with local sound speed  $c_{\rm s}(r)$
subject to gravitational
interactions with a planet. It includes the effects of pressure and gravity perpendicular to the disc plane that vary
along the eccentric
orbits.  The theory includes the effects of vertical oscillations of the eccentric disc resulting from a lack of vertical hydrostatic equilibrium that in turn leads to a prograde contribution to the precession of the disc  \citep{Ogilvie08}.
We drop the planet-disc potential terms $\Phi_{\rm pd}$ and add
the KL terms given by Equation (\ref{E1}), together with $d \Omega/dt$ given by Equation (\ref{dOmdt0}). 

We consider a disc that extends from inner radius  $r_{\rm in}$ to outer radius $r_{\rm out}$ and has surface density $\Sigma(r)$.
The evolution equation is given by
\begin{equation}
\begin{split}
 {\rm i}  \partial_r \left ( b_1(r) \partial_r \left (\frac{E(r,t)}{c_{\rm s}^2} \right ) \right )+   {\rm i}  b_2(r) E(r,t) \\  + {\rm i}  b_3(r)  E^*(r,t)  
  = J(r) \partial_t E(r,t). \label{E_eqiso} 
 \end{split}
\end{equation}
Quantity  $J(r)$ is the disc angular momentum per unit radius divided by $\pi$ and is given by
\begin{equation}
J(r) = 2 r^3 n(r) \Sigma(r),
\end{equation}
where $n(r)$ is the Keplerian orbital frequency 
about  mass $M_{1}$ given by
$n(r) = \sqrt{G M_1/r^3}.$
Functions $b_i(r)$ are given by
\begin{eqnarray}
b_1(r) &=& \Sigma(r) c_{\rm s}^4(r)  r^3, \\
b_2(r) &=&    3  \Sigma(r) \frac{d  (  r^2  c_{\rm s}^2(r))}{dr}+\frac{d  (\Sigma(r)   c_{\rm s}^2(r))}{dr} r^2  \label{b2}\\&+&  J(r)  A(r), \nonumber \\
b_3(r) &=&  J(r)  B(r),
\label{ab}
\end{eqnarray}
where
$c_{\rm s}(r)$ is the local disc sound speed, $\Sigma(r)$ is the disc surface density distribution, and
$A(r)$ and $B(r)$ are gravitational terms due to the binary for small $e$ that are associated with Kozai--Lidov oscillations in an initially slightly eccentric disc and are given by
\begin{eqnarray}
A(r)&=& \frac{3 G M_2}{8 n(r) a_{\rm b}^3} (2 - 3 \sin^2{i}) - \frac{d \Omega}{dt} \cos{i},\\
\label{Ar}
B(r)&=& \frac{15 G M_2}{8 n(r) a_{\rm b}^3} \,  \sin^2{i},
\label{Br}
\end{eqnarray}
with 
\begin{eqnarray}
 \frac{d \Omega}{dt} &=& - \frac{3 G M_2}{4 a_{\rm b}^3} \,\, \frac{\int  r^3 \Sigma(r) dr}
 {\int   n(r) \Sigma(r)  r^3  dr}  \, \cos{i},
\label{dOmdt}
\end{eqnarray}
where the integrals in Equation (\ref{dOmdt}) extend over
$r$ from $r_{\rm in}$ to $r_{\rm out}$.
$A(r)$ is the binary-induced apsidal precession rate relative to the line of ascending nodes, $\partial_t \, \omega(r,t)$ 
as would occur for a ballistic
particle.

We adopt the inner and outer boundary conditions that
\begin{equation}
\partial_r E( r_{\rm in},t) = \partial_r E( r_{\rm out},t)=0.
\label{EBC}
\end{equation}
For a Keplerian disc, the divergence of the velocity is proportional to $\partial_r E$.
Consequently, this boundary condition is equivalent to requiring that the
Lagrangian pressure perturbations near the disc edges vanish.

We consider normal modal solutions to Equation (\ref{E_eqiso}).
The KL effect  introduces a nonanalytic term involving $E^*$.
We follow the method of solution given by equations (37)--(40) of \cite{Lubow00} that involves
a similar nonanalytic term but describes tilt, rather than eccentricity.
We consider modes of the form
\begin{equation}
E(r, t) = E_+(r) \exp{(\lambda \,  t)} + E_-(r) \exp{( \lambda^* \, t)},
\label{Epm}
\end{equation}
where $\lambda$ is the complex growth rate.
The modal equations are given by 
\begin{equation}
 E_+'' + c_1 E_+' + c_2 E_+ + c_3  E_-^* = -\lambda \, c_4 E_+
 \label{Ep}
\end{equation}
and
\begin{equation}
 E_-'' + c_1 E_-' + c_2 E_- + c_3  E_+^* = -\lambda^* \, c_4 E_-,
 \label{Em}
\end{equation}
where
\begin{eqnarray}
c_1(r) &=&  \frac{3}{r} + \frac{\Sigma'}{\Sigma}, \\
c_2(r) &=& \frac{6}{r^2}+\frac{2 A \, n}{T} +  \frac{T'}{r T} +  \frac{\Sigma'}{r \Sigma} - \frac{\Sigma'\, T'}{\Sigma\, T}- \frac{T''}{T},  \label{c2} \\
c_3(r) &=&  \frac{2 B \, n}{T} ,  \\
c_4(r) &=&  \frac{2 {\rm i}\, n}{T} ,  \ \
\label{cn}
\end{eqnarray}
where a prime denotes differentiation by $r$, $T=c_{\rm s}^2$ is proportional to the temperature, and $A$ and $B$ are given by Equations (\ref{Ar}) and (\ref{Br}).
We apply the following boundary conditions:
\begin{eqnarray}
 E_+'(r_{\rm out})&=&0, \label{EpoNBC} \\ 
 E_-'(r_{\rm out})&=&0, \label{EmoNBC} \\
 E_+'(r_{\rm in})&=&0, \label{EpiNBC}\\
 E_-'(r_{\rm in})&=&0, \label{EmiNBC}\\
 e(r_{\rm out })&=&|E(r_{\rm out},t=0)| \nonumber \\
 &=&|E_+(r_{\rm out}) + E_-(r_{\rm out})|=1. \label{ENorm}
\end{eqnarray}
The latter condition imposes a normalization constraint on $e$ that is arbitrary,
since the equations are linear. The values of $e(r)$ are then measured relative to $e(r_{\rm out})$.

\section{Description of Models} \label{sec:models}

We explore the properties of KL oscillations in a set of models
all of which involve an equal mass binary and have outer disc edge $r_{\rm out}$
set to $0.3 a_{\rm b}$. This disc outer radius is  slightly larger than the value of about $0.25 a_{\rm b}$ for a disc in an equal-mass binary assuming coplanar orbits \citep{Paczynski77}. A misaligned disc feels a weaker binary torque and thus the outer truncation radius is larger  \citep[e.g.][]{Lubow15,Nixon15,Miranda2015} and so we adopt a larger value.  
The other parameters are listed in Table 1,
where $p$ and $s$ are defined by $\Sigma(r) \propto r^{-p}$ and $T(r) = c^2_{\rm s} \propto r^{-s}$.
Models A1 and A2 involve flared discs  that have a disc aspect ratios $H/r \propto r^{1/8}$.
This case applies to a standard `active' accretion disc \citep{Lynden-Bell74}.
An active protostellar disc is typically dominated by accretional heating within about $1$ AU
from the central star, where it would follow such a temperature profile.
Models B1 and B2 involve discs with somewhat greater flaring, having disc aspect ratios $H/r \propto r^{1/4}$.
Such a level of flaring (or more) is expected in so-called `passive' protostellar discs, where the heating is dominated
by the contributions from the central star at all radii, or in active discs on scales greater than $\sim 1$ AU \citep{Chiang97}.
We explore a range of  disc aspect ratios at the disc outer edge $h_{\rm out}=H/r(r_{\rm out})$, typically between 0.02 and 0.15.

Observations of protostellar discs suggest that the  surface density parameter power law exponent $p$ is in the range of 0.5 to about 1, although there is considerable uncertainty  \citep{Williams11}. We adopt a value of 1.
Since we are interested in fairly wide binaries,
we apply two values for the disc inner radius that are small compared to the
binary separation. The orbital radii $\sim 10  R_\odot$  for the so-called hot Jupiter planets
suggests a crude estimate for the central hole size in the disc where migrating planets are trapped \citep{Lin96}.
 For a binary with separation $\sim 500$ AU and period $\sim 10^4$ yr,
and a disc with a $\sim 10$ solar radii inner hole, the inner disc radius $r_{\rm in} \sim 10^{-4} a_{\rm b}$.
Consequently, it is important to analyze the eccentricity behavior at such small relative radii. We consider two
values of inner radii with other parameters fixed in order to probe the sensitivity of the results to $r_{\rm in}$.

\begin{table}
 \caption{Model Parameters}
 \label{tab:ModelParams}
 \begin{tabular}{lccc}
  \hline
  Model & $p$ & $s$ & $r_{\rm in}$  \\
  \hline
  A1 & 1 & 0.75 & 0.01 \\
A2 & 1 & 0.75 & 0.0001 \\
  B1 & 1 & 0.5 & 0.01 \\
B2 & 1 & 0.5 & 0.0001 \\
    \hline
 \end{tabular}
\end{table}

\section{Time Evolution and Methods of Solution}
\label{sec:mos}

\begin{figure*}
\centering%
\includegraphics{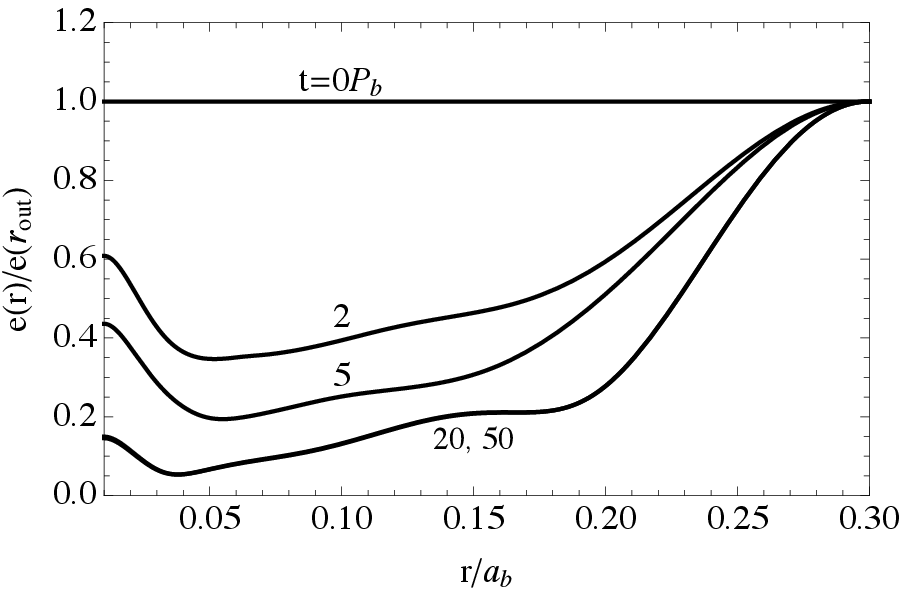}
\includegraphics{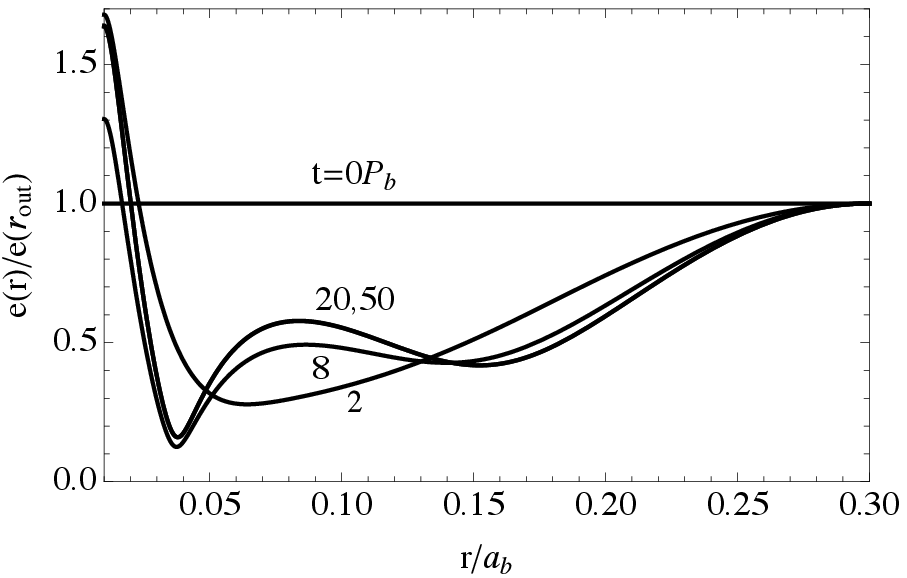}
\caption{Evolution of eccentricity for Model A1 with disc inclination $i= 45^\circ$ and $h_{\rm out}=0.03$ (upper plot) and $h_{\rm out}=0.05$ (lower plot).
$e(r)$ is plotted at various times, normalized
by its value at the disc outer edge.
The disc eccentricity is initially constant in radius.  By a time of $20  P_{\rm b}$ the eccentricity distribution
settles into an eigenmode in which it is nearly the same as at time $50  P_{\rm b}$.  }
\label{fig:timeint}
\end{figure*}
We determine the eccentricity evolution by applying similar methods used in \cite{Lubow10} with Mathematica.
We compute the complex eccentricity evolution
given by Equation (\ref{E_eqiso}) in space and time for  model A1 with the initial condition that
 $E(r,0)=1$ for $r_{\rm in} \le r  \le r_{\rm out}$ using the method of lines.
 The initial eccentricity satisfies boundary conditions (\ref{EBC}).
 The results for this model 
 with an initial tilt $i=45^{\circ}$ and disc aspect ratios at the disc outer edge of  $h_{\rm out}=0.03$ and 0.05 are plotted in Figure~\ref{fig:timeint}.
After less than 20 binary orbital periods $P_{\rm b}$, $E(r,t)/|E( r_{\rm out},t)|$ settles closely to an eigenfunction in which $\partial_t E/E$ is nearly constant
 in $r$ and is equal to the eigenvalue.
  As we will see later, there is more than one growing eigenmode in each case. The plotted distributions at $t=50 P_{\rm b}$ reflect the fastest growing eigenmode that emerges as dominant for each case.  
  The eccentricity distributions are quite different for the two cases. The final eccentricity distribution has a lower eccentricity at the inner edge than the outer edge for the $h_{\rm out}=0.03$ case, while just the opposite occurs for the $h_{\rm out}=0.05$ case.   
  
 We numerically determine eigenfunction $E(r)$ and eigenvalue $\lambda$ in Equations (\ref{Epm})--(\ref{Em}) 
 by a shooting method. 
Equations  (\ref{Ep}) and  (\ref{Em}) are second order in space and
are integrated inward in radius from $r=r_{\rm out}$ to the inner boundary $r=r_{\rm in}$ by using the starting conditions for the radial derivatives given by
Equation (\ref{EpoNBC}) and  (\ref{EmoNBC}) at the outer radius $r_{\rm out}$.
The normalization condition (\ref{ENorm}) is replaced by 
\begin{eqnarray}
E_+(r_{\rm out})&=& w \exp{({\rm i} \, \omega_{\rm out})}, \label{EpoNorm}\\
  E_-(r_{\rm out})&=& (1-w) \exp{({\rm i} \, \omega_{\rm out})}, \label{EmoNorm}
 \end{eqnarray}
for  the real weighting factor $w$ and phase $\omega_{\rm out}$ at the disc outer edge. 
 The complex inner boundary conditions (\ref{EpiNBC}) and  (\ref{EmiNBC}) are equivalent to four
 real conditions. These are satisfied by adjusting the four real parameters 
 $Re(\lambda)$,  $Im(\lambda)$, $\omega_{\rm out}$,
 and $w$. 
Once an eigenmode is determined for a model, parameter changes from that eigenmode, such as in the disc inclination $i$, can then be made incrementally and iteratively with rapid convergence.
Consequently, it is more efficient to solve the eigenvalue problem, rather than the initial value problem.

Although we do not explicitly consider cases with $ i > 90^{\circ}$, the growth rates are invariant when $i$ is replaced by $180^{\circ} - i$.

\section{Role of Resonances}

The KL instability can be understood as a consequence of a resonance in which 
\begin{equation}
 \frac{d \omega}{dt}=0,  \label{res1}
\end{equation}
 or equivalently the matching of apsidal and nodal precession rates
\begin{equation}
\frac{d \varpi}{dt} = \frac{d\Omega}{dt}, \label{res}
\end{equation}
where $\varpi$ is the  longitude of the periapsis.
A test particle is subject to  perturbations that are solely due to the gravitational forces of the companion.
From Laplace's equation, it follows that near the binary orbital plane 
\begin{equation}
2 n^2 - \nu^2 - \kappa^2 =0,
\end{equation}
where $\nu$ is the vertical oscillation frequency and $\kappa$ is the epicyclic oscillation 
frequency of a test particle.
Using this identity and that  $\nu  \simeq \kappa \simeq n$, it follows that 
\begin{equation}
\frac{d \varpi}{dt} \simeq - \frac{d \Omega}{ dt} \simeq \frac{3 G M_2}{4 n a_{\rm b}^3}, \label{respartco}
\end{equation}
for $r \ll a_{\rm b}$, where $d \varpi/dt =  n-\kappa$ and $d \Omega/dt =  n- \nu.$
Consequently, resonance condition (\ref{res}) is impossible to satisfy for a 
nearly coplanar test particle orbit where $i \simeq 0$.

In the case of a test particle, 
we have from Equation (\ref{domegadt}) that
\begin{eqnarray}
\frac{d \varpi}{dt}  &=& \frac{3 G M_2}{4 n a_{\rm b}^3} (2 - \cos{i} -5 \sin^2{i} \sin^2{\omega}), \label{dvarpidt}\\
  \frac{d \Omega}{ dt} &=& -\frac{3 G M_2}{4 n a_{\rm b}^3} \cos{i}. \label{dOMdt}
\end{eqnarray}
Notice that for small $i$, these equations recover Equation (\ref{respartco}).
With increasing inclination $i \leq 90^{\circ}$, the apsidal precession rate given by Equation (\ref{dvarpidt}) can become negative, while the nodal
precession rate in Equation (\ref{dOMdt}) remains negative. Resonance is first  possible when $\omega =90^{\circ}$ ($\omega$ is constant in time by Equation (\ref{res1})), in order to make 
the apsidal rate as negative as possible. Applying this $\omega$ value and Equations (\ref{dvarpidt}) and (\ref{dOMdt}) to resonance condition (\ref{res}), we then recover the KL instability 
condition for a test particle given at the end of Section \ref{sec:partev} that $\cos^2{i} < 3/5$.
We see that KL instability is tied to satisfying the resonance condition.

The effects of pressure on a fluid disc modify the apsidal precession rate and can act to make it negative, even for a coplanar disc.
Consequently, as we will see, the coplanar KL resonance condition can be  satisfied for certain disc parameters.
There are several (formally infinitely many) values of disc aspect ratios for which the coplanar KL condition is satisfied.
These coplanar resonances play a key role in delineating the various branches of instability, as we will see in the next section. 

In this paper we are concerned with KL instability. Consequently we describe instability that is associated with
the condition that $d \omega/dt=0$
or 
\begin{equation}
Im(\lambda) =0 \label{Imlam}
\end{equation}
in Equation (\ref{Epm}).
Other forms of eccentric disc instability that do not satisfy this condition are possible 
\cite[e.g.,][]{Lubow91}.

\section{Marginal Stability}

We determine the conditions for marginal KL stability 
of the  disc models described in Section \ref{sec:models}.
Marginally stable  KL modes  have $Re(\lambda)=0^+$ and $Im(\lambda)=0$ by Equation (\ref{Imlam}). These
modes are stable, but are very close to instability.
We determine some  properties of the marginally stable modes by using Equation (\ref{E_eqiso}) with the RHS set to zero, since $\partial_t E=0$.
Expressing $E = e_{\rm x} + i e_{\rm y}$ for real $e_{\rm x}$ and $e_{\rm y}$ and taking the real and imaginary parts of that equation, we find that there are modes
that satisfy
\begin{eqnarray}
 \frac{d}{dr} \left ( b_1 \partial_r \left (\frac{ e_{\rm x} }{c_{\rm s}^2} \right ) \right ) +    (b_2  +  b_3)  e_{\rm x} &=&0,  \label{exmarginal} \\
  e_{\rm y} &=&0
\end{eqnarray}
with normalization condition
\begin{equation}
e_{\rm x}(r_{\rm out})=1,
\label{exnorm}
\end{equation}
and other modes that satisfy
\begin{eqnarray}
 e_{\rm x} &=&0,  \label{eymarginal} \\
 \frac{d}{dr} \left ( b_1 \partial_r \left (\frac{ e_{\rm y} }{c_{\rm s}^2} \right ) \right ) +    (b_2  -  b_3)  e_{\rm y} &=&0
\end{eqnarray}
with normalization condition
\begin{equation}
e_{\rm y}(r_{\rm out})=1.
\label{eynorm}
\end{equation}

The boundary condition given by Equation (\ref{EBC}) implies that 
\begin{equation}
\frac{d e_j}{dr}(r_{\rm in}) = \frac{d  e_j}{dr}(r_{\rm out})=0.
\label{exyBC}
\end{equation}
for $j={\rm x}, {\rm y}$.

The marginal modes are therefore untwisted and have a constant phase of either $\omega = (0^{\circ},180^{\circ})$
 or $\omega = (90^{\circ},270^{\circ})$.
We note that the same phase results occur for 2D discs and adiabatic discs
using appropriate eccentricity evolution equations given by
\cite{Teyssandier16}.
Equations (\ref{exmarginal})--(\ref{exnorm}) or Equations (\ref{eymarginal})--(\ref{eynorm}), together with Equation (\ref{exyBC}) are regarded as an eigenvalue 
problem for the critical tilt angle for the marginally stable mode  with $i=i_{\rm crit}$.

We consider all parameters fixed, except the disc aspect ratio at the disc outer edge 
$h_{\rm out}=H/r(r_{\rm out})$ and the inclination
$i$. We determine the critical angle $i_{\rm crit}$ as a function of $h_{\rm out}$ for all marginal modes. 
To determine the critical angles for modes with $\omega=(0^{\circ},180^{\circ})$, 
we start with a given value of $h_{\rm out}$ and integrate Equation (\ref{exmarginal}) inward with outer boundary condition (\ref{exyBC}) and normalization
condition (\ref{exnorm}). We determine the values of
$i=i_{\rm crit}$ such that the inner boundary condition
(\ref{exyBC}) is satisfied. That is, by integrating inwards for several assumed $i$ values, we determine the values of 
$d e_{\rm x}/dr (r_{\rm in})$.
We find the $i$ values near where $d e_{\rm x}/dr (r_{\rm in})$
changes sign and refine the search  for $i_{\rm crit}$ with a local
root finder. With this technique, we believe that we determine all marginal modes at a given $h_{\rm out}$.
The same process is applied to modes with $\omega=(90^{\circ},270^{\circ})$.

 In general there exists more than one $i_{\rm crit}$  for the same $h_{\rm out}$ corresponding to different modes with a different number of radial nodes in the eigenfunctions $e(r)$. 
We follow these marginal modes to higher and lower values of $h_{\rm out}$ and obtain $i_{\rm crit}(h_{\rm out})$.

\begin{figure*}
\centering%
\includegraphics{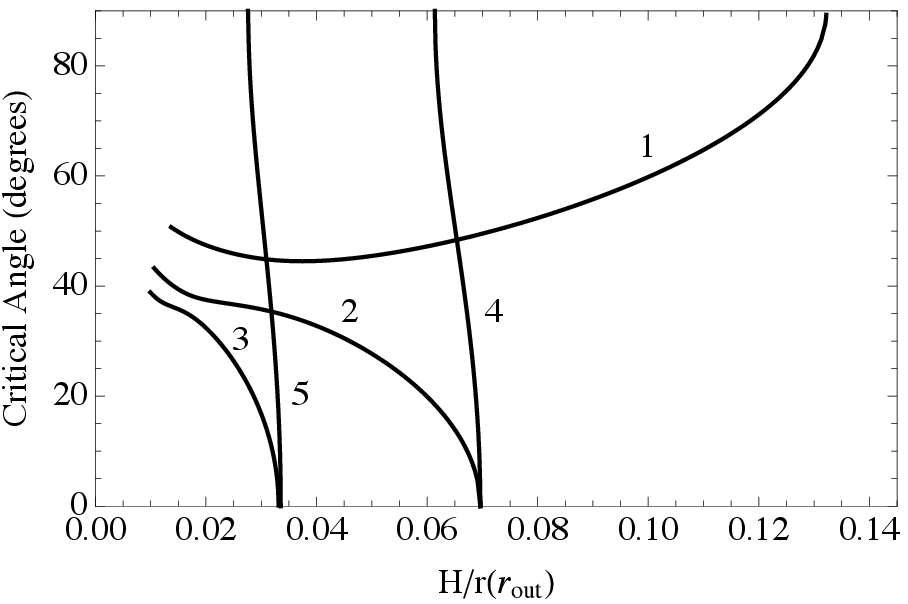}
\includegraphics{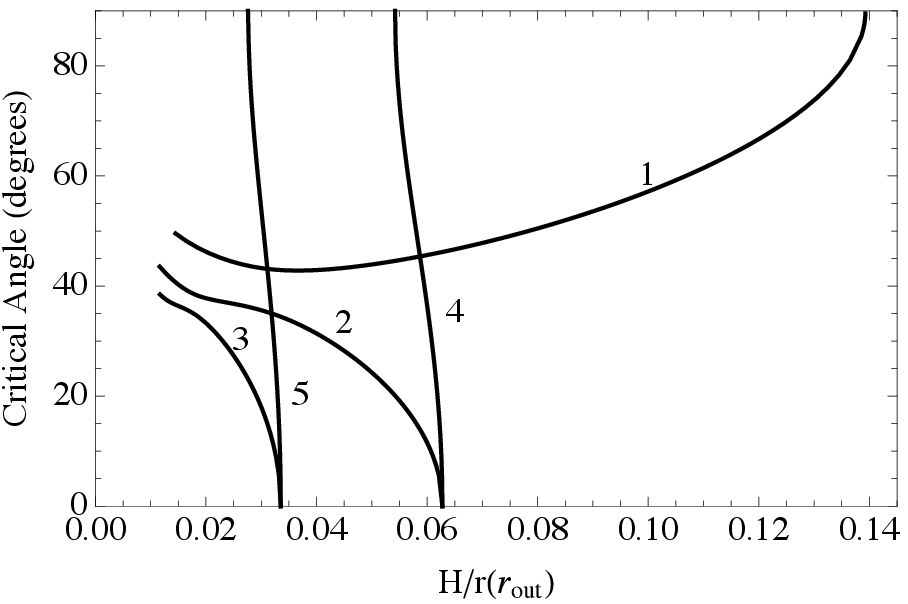}
\caption{Critical disc inclination $i_{\rm crit}$ as a function of disc aspect ratio $H/r(r_{\rm out})$ for marginally stable modes
of disc model A2 (top panel) and model B2 (bottom panel). The branches are labeled by integers.}
\label{fig:marginalA2B2}
\end{figure*}

Figure~\ref{fig:marginalA2B2} plots the inclinations $i_{\rm crit}$ for the five marginal modes with the highest $h_{\rm out}$ values for models A2 and B2 over $0.02 \la h_{\rm out} \la 0.15$. The  branch numbers are not directly related across models. We later determine this relationship.
For both models, branches 1, 2, and 3 have phase $\omega_{\rm out}=90^{\circ}$, while the nearly vertical branches 4 and 5
have phase  $\omega_{\rm out}=0$ and cross the other branches.
The values of $i_{\rm crit}(h_{\rm out})$ along branch 4 are $\sim 3\, i_{\rm crit} (h_{\rm out})$ along branch 2, until branch 4 reaches $i=90^{\circ}$.
The same relationship exists between branches 3 and 5.
Notice that all branches except for branch 1 extend
to coplanarity, $i=0$. 
 For larger $h_{\rm out}$, the marginal stability of branch 1 extends to $90^{\circ}$.
More branches occur at smaller $h_{\rm out}$ than are plotted here.
 There is an another marginally stable branch for model B2 with $i=90^{\circ}$ at $h_{\rm out} \simeq 0.025$ that resembles
branch 1 that we do not include in the plot.
 Notice that the number of branches with phase $\omega_{\rm out}=90^{\circ}$ (branches 1, 2, and 3) at the same $h_{\rm out}$ value increases with decreasing $h_{\rm out}$.
The marginal mode branches for models A2 and B2 plotted in Figure~\ref{fig:marginalA2B2} have a very similar
topology, but are shifted somewhat relative to each other.

\begin{figure*}
\centering%
\includegraphics{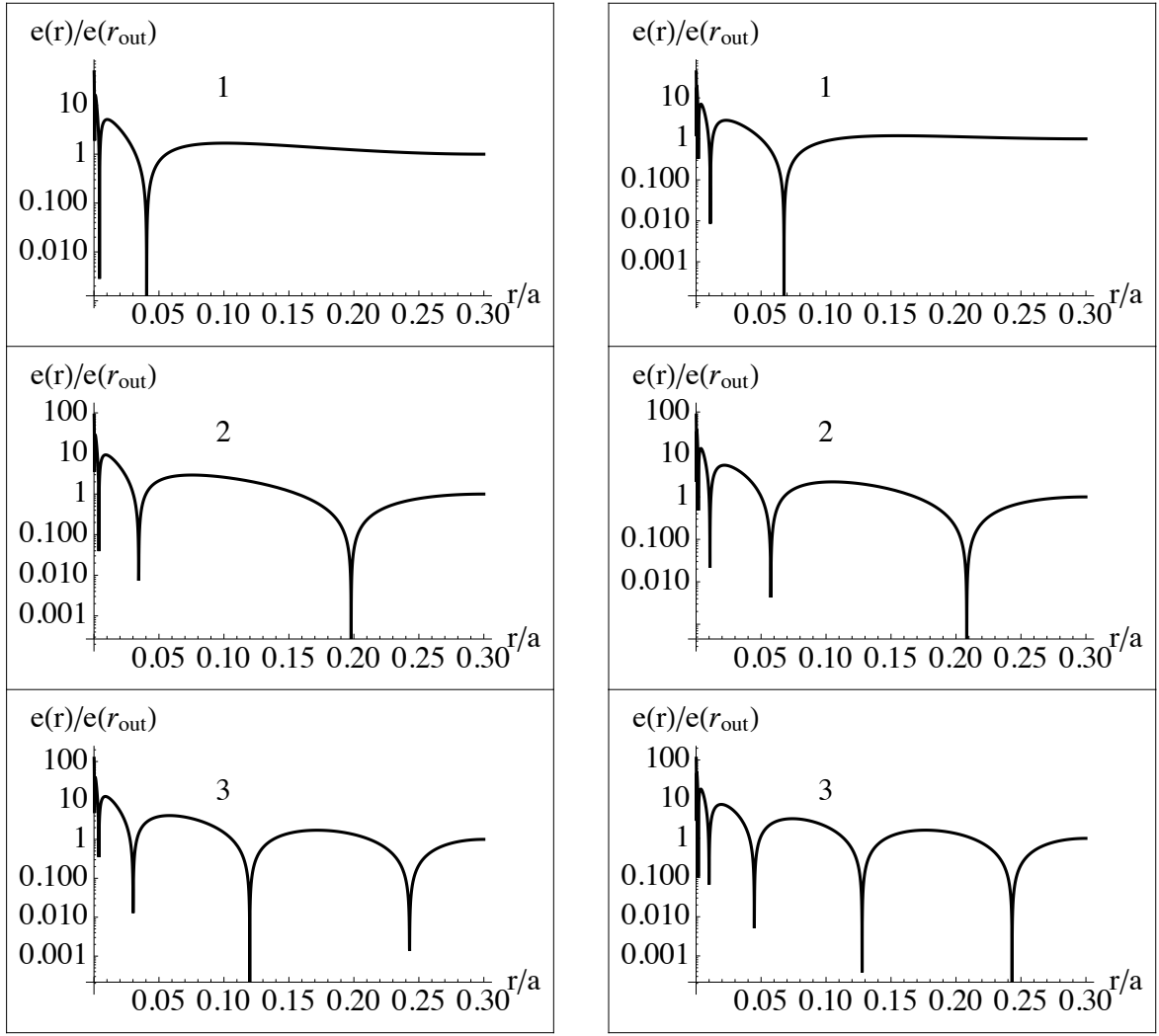}
\caption{Marginally stable  eigenmodes for $i_{\rm crit}=90^{\circ}$ for branch 1, and  $i_{\rm crit}=0^{\circ}$  for branches 2 and 3 in Figure \ref{fig:marginalA2B2} for models A2 (left panel) and B2 (right panel) .
 }
\label{fig:marginalmodesA2B2}
\end{figure*}

Figure~\ref{fig:marginalmodesA2B2} plots the eigenfunctions for branches 1, 2, and 3 of models A2 and B2.  The vertical scale is logarithmic and the dips in the plots correspond to nodes. The number of nodes in the region of the disc with $r \ga 0.02 a_{\rm b}$ is seen to increase with the branch number.
The modes along a given branch of a model have the same number of nodes.
The behaviour at small $r$  is hard to discern in this plot, but will be described in Section \ref{sec:eigenprops}. 
The eigenfunctions for branches 4 and 5 are very similar to those for branches  2 and 3, respectively, but are shifted in phase.

\begin{figure*}
\centering%
\includegraphics{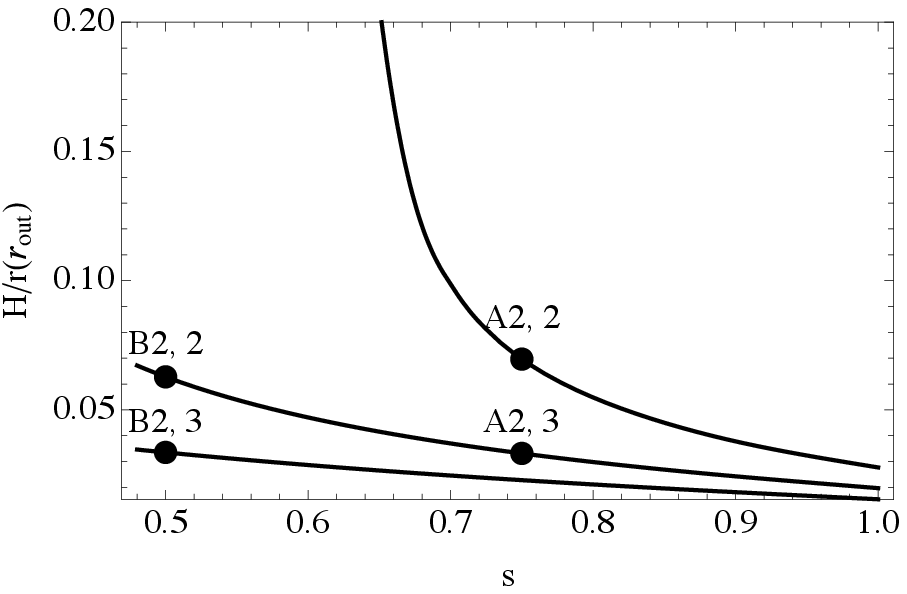}
\caption{ $H/r(r_{\rm out})$  for some  coplanar KL resonances  as a function of the temperature
exponent $s$ for a fixed density distribution with exponent $p=1$. The points are labeled by model and branch number within the model.
 }
\label{fig:tracks}
\end{figure*}

\begin{figure*}
\centering%
\includegraphics{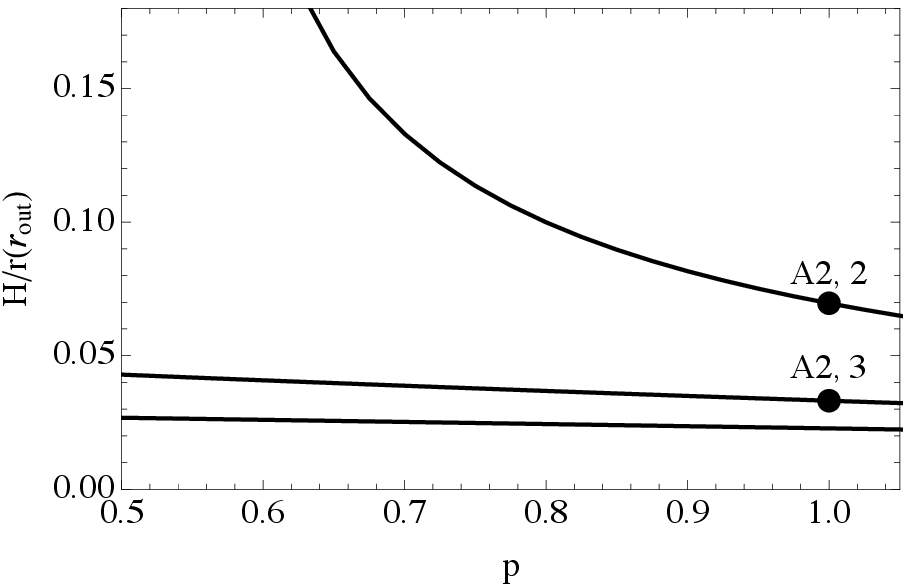}
\caption{ $H/r(r_{\rm out})$  for some  coplanar KL resonances as a function of the density
exponent $p$ for a fixed temperature distribution with exponent $s=0.75$.  The points are labeled by model and branch number within the model.
 }
\label{fig:trackp}
\end{figure*}

In 
Figure~\ref{fig:tracks} we plot the $h_{\rm out}$ values for some of the coplanar KL resonances 
as a function of the temperature
exponent $s$ for a fixed density  exponent $p=1$.  Some points are labeled by model and branch number.
For example, $h_{\rm out}=H/r(r_{\rm out}) \simeq 0.07$ for the point on the plot that is labeled  for branch 2 of model A2 corresponds 
to the $h_{\rm out}$ value at zero inclination in the top panel of Figure~\ref{fig:marginalA2B2} for branch 2.
The effect of increasing $s$ is to shift the resonances towards lower $h_{\rm out}$. The direction of this shift is in the sense of reducing changes to the pressure gradient, since an increase in $s$ raises the pressure gradient at fixed sound speed, while the decrease in $h_{\rm out}$ lowers the sound speed.
We see from the plot that branch 3 of model A2
lies on the same curve as branch 2 of model B2. Consequently, they correspond to the same mode structure, i.e., have same number of nodes in the eigenfunction.  
We then expect that the branch numbers for the same mode structure differ by unity across the two models. Therefore, branch 2 of model A2
corresponds to branch 1 of model B2. The mode structure of branch 1 of model A2 is not present in model B2.
The mode structure for branch 3 of model B2 exists at smaller $h_{\rm out}$ than is plotted for model A2 in the top panel of Figure~\ref{fig:marginalA2B2}.
In Figure~\ref{fig:trackp} we plot $h_{\rm out}$ for some of the coplanar KL resonances as a function of the density
exponent $p$ for a fixed temperature exponent $s=0.75$.  
The effect of increasing the exponent is again to shift the resonances towards lower $h_{\rm out}$.

\section{Zones of Instability}

We apply the marginally stable state information to determine
the parameter ranges for KL instability (requiring $Im(\lambda)=0$). We expect that modes 
will be growing in time for inclinations $i$ above the marginal stability
curves in Figure~\ref{fig:marginalA2B2}.   

We consider continuous sequences of modes (eigenfunctions and eigenvalues) for some model that start on marginally stable branch 1, 2, or 3
and extend to larger $i$ at fixed $h_{\rm out}$ (vertically upward from $i_{\rm crit} (h_{\rm out})$ on Figure~\ref{fig:marginalA2B2}). We consider these modes as functions of parameters, that we denote as $M(b, h_{\rm out}, i)$, where $b$ is the starting branch number 
1, 2, or 3 of the mode sequence.
From brevity, we sometimes refer
to such a  mode as being on an extension
of a particular branch. For example, a mode  that  is obtained from  a continuous sequence of modes  at fixed $h_{\rm out}$ that start  on branch 1  is referred to as 
lying on an extension of branch 1.
For a given $h_{\rm out}$ and inclination $i$, any branches $b$ with 
$i > i_{\rm crit}$ may  provide  unstable modes that lie in their extensions. In general, there can be multiple unstable
modes that are extensions of different branches.

\begin{figure*}
\centering%
\includegraphics[width=12cm]{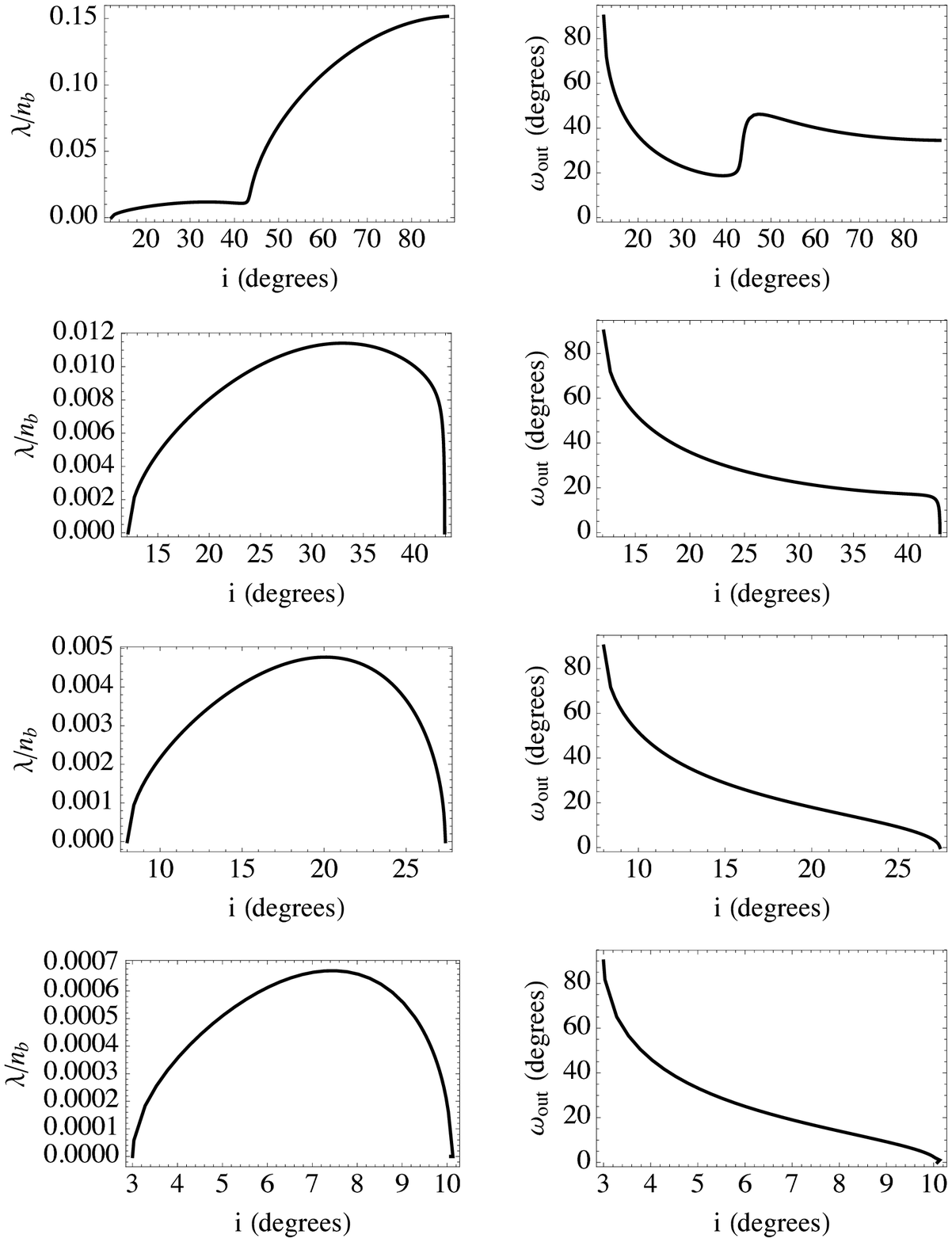}
\caption{Left column: Growth rates as a function of inclination for   $H/r(r_{\rm out})=0.06607, 0.06616, 0.06812$ and $0.06941$
 (top to bottom) for modes
that start at marginal stability on branch  2 of model A2 at $i=12.25^{\circ}, 12.1^{\circ}, 8^{\circ},$ and $3^{\circ},$ 
 respectively, and continuously
 extend to larger tilt angles $i$.
Right column: Phase $\omega_{\rm out}$ as a function of inclination for the adjacent case
in the left column of the same row.
 }
\label{fig:marginalmodes24}
\end{figure*}

Figure~\ref{fig:marginalmodes24} plots the growth rates and phases as a function of inclination
for modes that lie on extensions  of branch 2 for model A2, that is $M(2, h_{\rm out}, i)$.
These cases have $h_{\rm out}$ values of
 0.06607, 0.06616, 0.06812, and 0.06941. 
 The peak growth rates increase with starting angle and decrease with $ h_{\rm out}$.
For the bottom three rows in the figure, the plots begin with phase 
$\omega_{\rm out}=90^{\circ}$ and end with  $\omega_{\rm out}=0^{\circ}$.
Recall that modes on branches 2 and 4 have constant $\omega$ with  $\omega_{\rm out}=90^{\circ}$ and $\omega_{\rm out}=0^{\circ}$, respectively.
Instability terminates with increasing $i$ on branch 4  with $\omega_{\rm out}=0^{\circ}$ at approximately 3 times the starting inclination angle on branch 2 for the lower three cases.
The absolute values of the derivatives at the endpoints of the curves in the left and right panels increase with starting angle and reach
a near infinite absolute value at  $i \simeq 43^{\circ}$ on the second panel from the top. For a slightly smaller aspect ratio, $h_{\rm out}=0.06607$ there is an
abrupt change of behaviour in which the terminating angle is no longer limited by branch 4 and is plotted
to $90^{\circ}$. 

\begin{figure*}
\centering%
\includegraphics{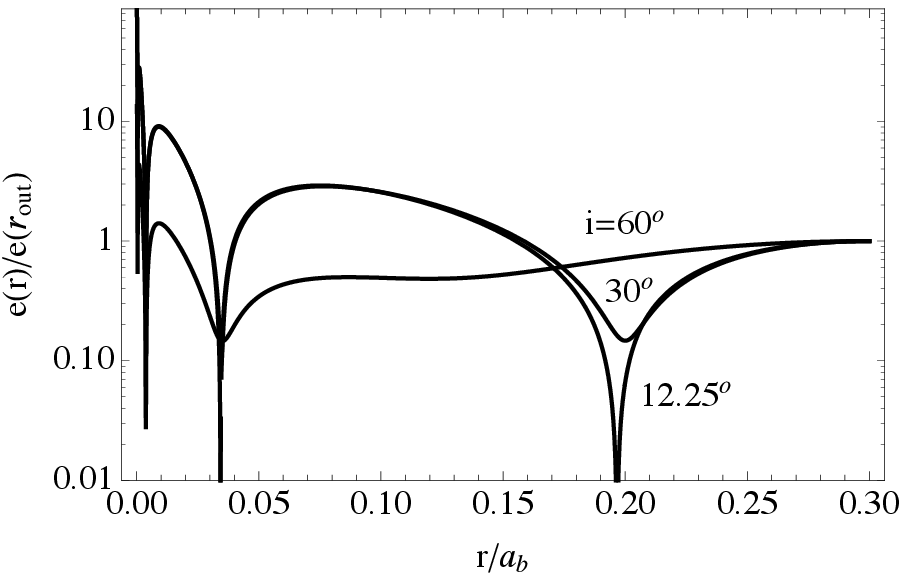}
\caption{Log plots of normalized eccentricity for model A2  with  $H/r(r_{\rm out})=0.06607$ (top row of Figure~\ref{fig:marginalmodes24})
that lie on an  extension of  branch 2 to i=$12.25^{\circ}, 30^{\circ}$, 
and $60^{\circ}.$
 }
\label{fig:modes0p06607}
\end{figure*}

The modes along a marginally stable branch have the same mode structure, i.e., the same number of nodes. However, modes that extend from a marginal mode branch to higher inclinations can undergo major structural changes.
Figure~\ref{fig:modes0p06607} plots the eigenfunctions  for the model
in the top  row of  Figure~\ref{fig:marginalmodes24} for three different tilt angles.
These modes lie on extensions of branch 2 of model A2 that correspond to  $M(2,0.06607, 12.5^{\circ})$,  $M(2,0.06607, 30^{\circ})$,
and $M(2,0.06607, 60^{\circ})$ in the  notation defined above. The mode at $i=12.5^{\circ}$
is marginally stable and lies on branch 2. Its eigenfunction resembles the branch 2 eigenfunction for a different $h_{\rm out}$ that is shown in the  middle  left panel of Figure~\ref{fig:marginalmodesA2B2}.
However, at larger tilt angles, the outer node disappears and the eigenfunction more closely resembles a branch 1 marginal mode of model A2, such as the one shown in the upper left of  Figure~\ref{fig:marginalmodesA2B2}.
Notice also that the knee in the growth rate curve shown in Figure~\ref{fig:marginalmodes24} occurs
at about $i \simeq 45^{\circ}$ that is roughly at the critical angle for branch 1,
$i_{\rm crit}(0.06607) \simeq 48^{\circ}$. Above this knee, the growth rate climbs substantially
and the eigenfunctions have a simpler structure that are similar to a branch 1 mode.

The abrupt change in behaviour across the top two rows of Figure~\ref{fig:marginalmodes24}
in going from  $h_{\rm out}= 0.06607$ to 0.06616
is likely related to the crossing of branch 1 with branch 4 in Figure~\ref{fig:marginalA2B2}.
The  models plotted in the  lower three rows of  Figure~\ref{fig:marginalmodes24}
have a positive growth rate over a much more limited  range of angles than the top row.
The reason is that these modes are terminated at  angles where they  intersect with  branch 4
at fixed $h_{\rm out}$ in proceeding vertically upward from branch 2 to branch 4 in Figure~\ref{fig:marginalA2B2}. 

\begin{figure*}
\centering%
\includegraphics{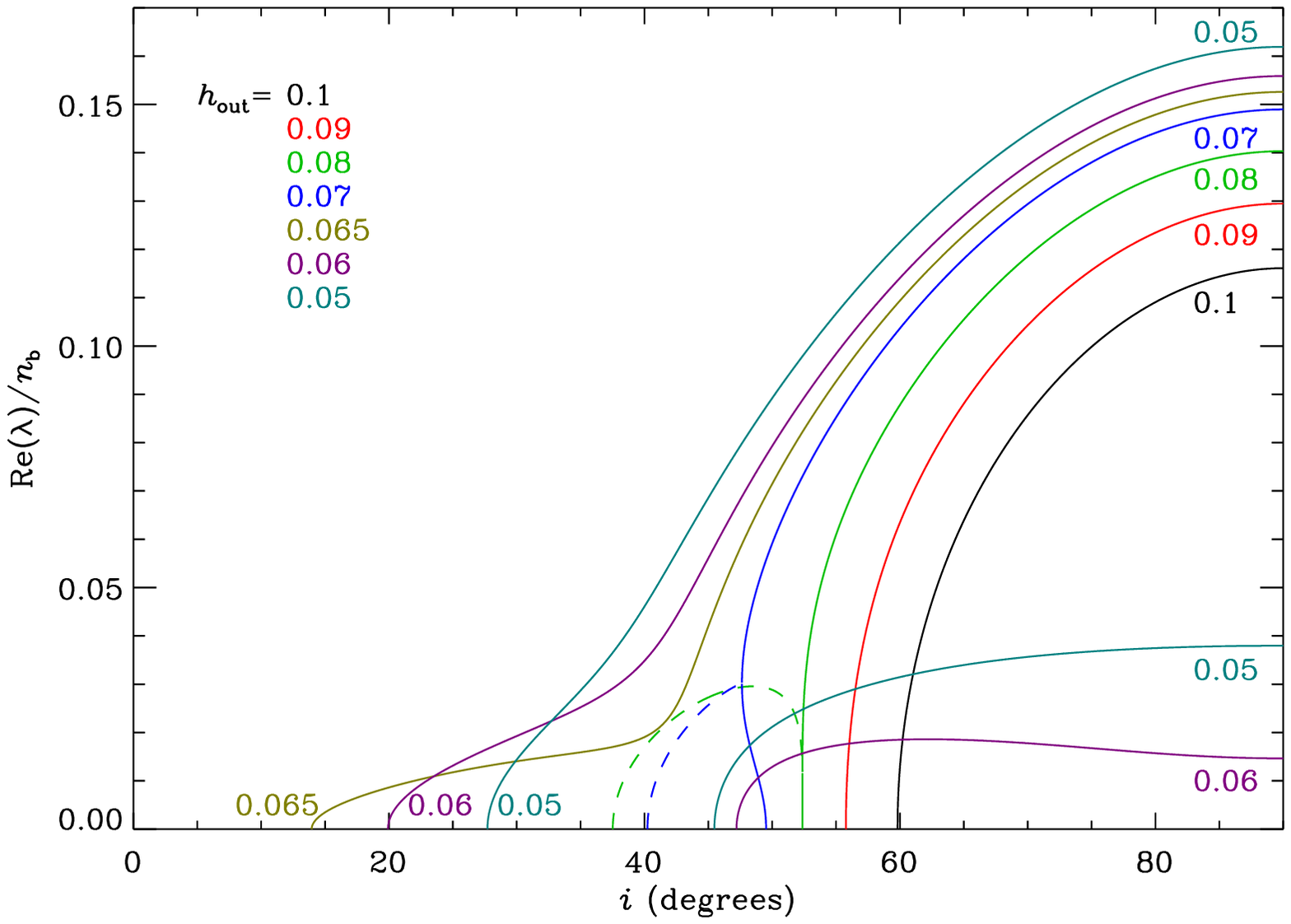}
\caption{Growth rates as a function of inclination for some sequences of modes at various values of $H/r(r_{\rm out})$ involving model A2.
The two dashed lines indicate  sequences of modes with complex growth rates that terminate at zero growth
rate to  marginally stable modes with nonzero precession frequencies.
 }
\label{fig:modelA2.eps}
\end{figure*}

\begin{figure*}
\centering%
\includegraphics{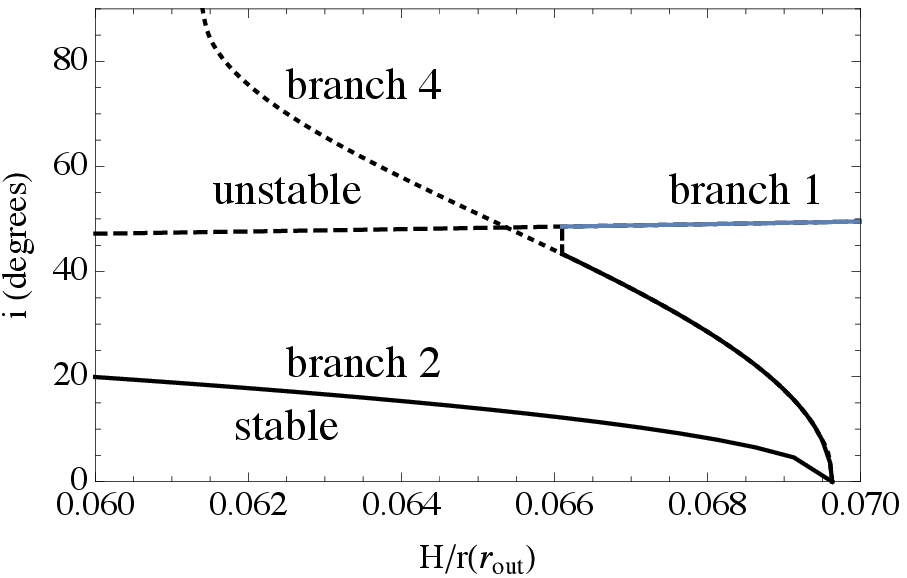}
\caption{Region of KL instability close to branch 4.
Unstable modes cover the region above the solid line for branch 2 and below the solid
line for branch 4. Unstable modes cover the region above the solid line for branch 1. Unstable modes also exist 
between the solid lines for branches 1 and 4. 
This region contains modes with complex growth rates that we
do not fully explore in this paper.
 }
\label{fig:mapn2}
\end{figure*}

 Figure \ref{fig:modelA2.eps} plots the growth rates for some sequences of modes involving model A2. 
For $h_{\rm out} = 0.05, 0.06$ and $0.065$, the sequences begin on branch 2, while for
 $h_{\rm out} = 0.07, 0.08, 0.09$ and 0.1 the sequences begin on branch 1. (In the case $h_{\rm out} = 0.065$ there is a short second sequence of unstable modes between branches 4 and 1, near $i=50^{\circ}$ that we do not plot in this figure.)
 In most cases, the growth rate increases monotonically with tilt angle above the critical tilt for marginal stability.
 The case of $h_{\rm out} = 0.07$, which is near the $h_{\rm out}$ value for the crossing of branches 1 and 4,
 is somewhat different.
The plot in that case
shows that unstable modes first extend below branch 1 before rising to higher inclinations with higher growth rates.
The sequence joins to another sequence of unstable modes 
with complex growth rates (indicated by the dashed line)  that extends to lower inclination. This sequence terminates near $i=40^{\circ}$ 
at a marginally stable mode that has a nonzero precession frequency. A similar behaviour occurs for $h_{\rm out} = 0.08$.
 
 A closeup of the region near branch 4 is shown in Figure~\ref{fig:mapn2}. 
  Branch 4  is shown as the dotted line that overlaps  with the solid line on the lower right.
The solid line for branch 4 is an upper boundary of the region 
of instability above branch 2 from $h_{\rm out} \simeq 0.0665$ to 0.0696. 
The region at low inclinations below branch 2 is stable. Branch 1 is plotted by a dashed line on the left  and the solid line on the right
of the dashed short vertical line.
 In addition, the region between the solid lines for branches 1 and 4 also contains unstable modes. Over the range of the solid line for branch 1 in this figure, sequences of unstable modes  extend below branch 1, as discussed above for $h_{\rm out} = 0.07$. Such sequences include a complex extension that is terminated by a marginal mode that we have not plotted because it has a non-zero precession frequency.
 We do not pursue a detailed analysis of the
stability of modes in this region in this paper. 
 The triangular region in the figure is an unstable zone below branch 1 that also includes a complex extension that is terminated by a marginal mode that has $Im(\lambda)\ne0$. It also includes
 the some of unstable modes from the mode sequence plotted top row in Figure~\ref{fig:marginalmodes24}.

Branch 4 extends from $h_{\rm out}=0.0614$ at $i=90^{\circ}$ to $h_{\rm out}=0.0696$ at $i=0^{\circ}$. 
The top row of Figure~\ref{fig:marginalmodes24} is for $h_{\rm out}= 0.06607$ that lies
within the span of $h_{\rm out}$ values of branch 4. Yet, the range of angles over which instability occurs
is not terminated by branch 4.
The abrupt change in behaviour across the top two rows of Figure~\ref{fig:marginalmodes24}
in going from  $h_{\rm out}= 0.06607$ to 0.06615
is likely related to the crossing of branch 1 with branch 4 in Figures~\ref{fig:marginalA2B2} and \ref{fig:mapn2}.

\begin{figure*}
\centering%
\includegraphics[width=10cm]{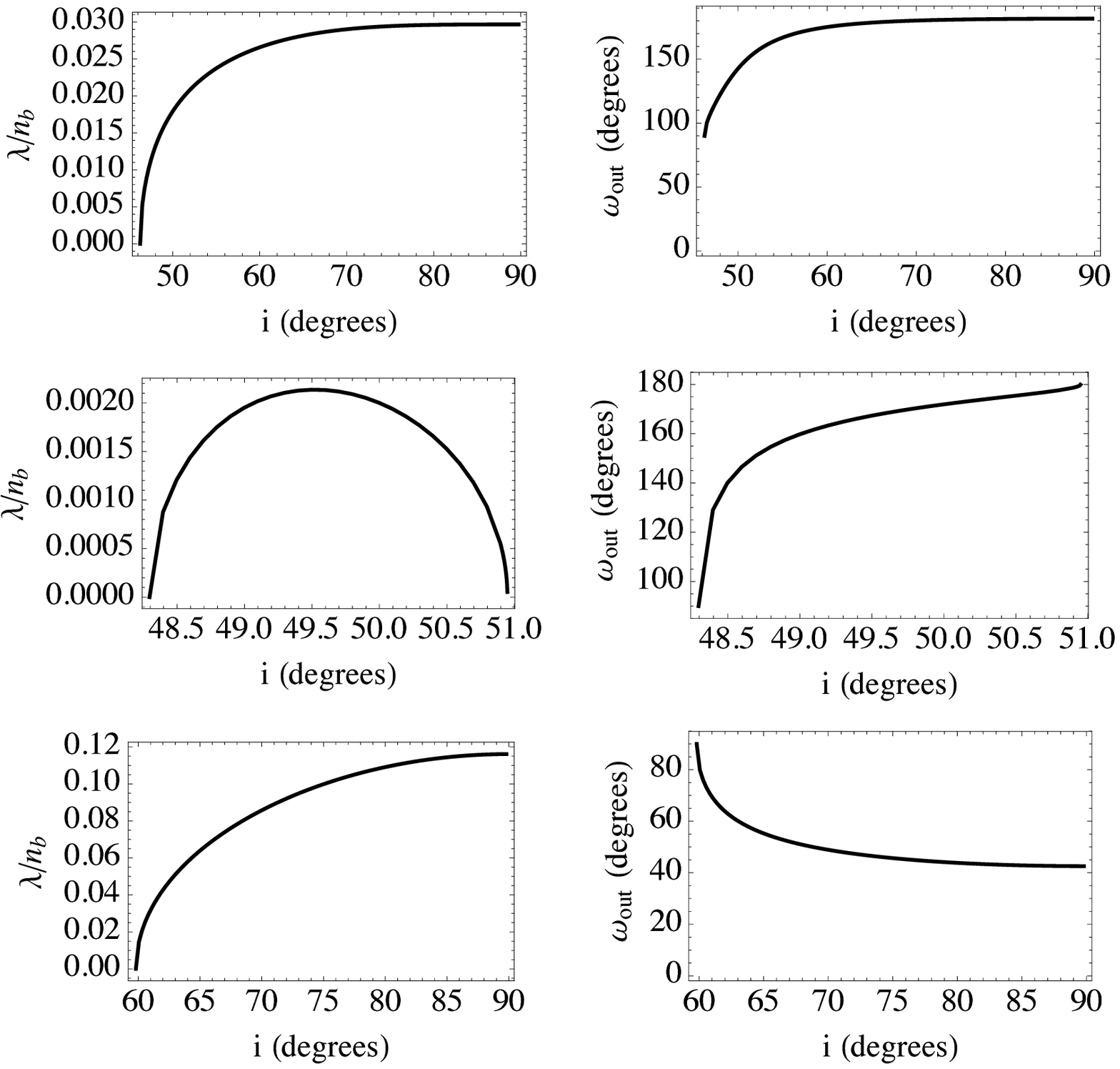}
\caption{Left column: Growth rates as a function of inclination for modes with $h_{\rm out}=0.055, 0.065,$ and 
$0.10$ from top to bottom panels respectively
on branch 1 of model A2. 
Right column: Phase $\omega_{\rm out}$ as a function of inclination for the adjacent case
in the left column.
 }
\label{fig:modesn1}
\end{figure*}

As branch 4 reaches close to branch 1 from lower $i$, its influence on modes that are an extension
of branch 2, that is 
$M(2, h_{\rm out}, i)$,  abruptly stops and shifts to modes that are on an extension of branch 1,
$M(1, h_{\rm out}, i)$.
This effect on modes that are an extension of branch 1 is seen in Figure~\ref{fig:modesn1}.  
Notice that the smallest value $h_{\rm out}=0.055$ that is plotted 
(top row in this figure) is smaller than of the minimum
$h_{\rm out}$ value spanned by branch 4, while the largest value  (top row in this figure) has  $h_{\rm out}=0.10$
that lies beyond the largest $h_{\rm out}$ value spanned by branch 4.
In both cases, the positive growth rates extend from the critical value of $i_{\rm crit}$ at marginal stability plotted
in Figure~\ref{fig:marginalA2B2} to $90^{\circ}$. However, the intermediate $h_{\rm out}=0.065$
lies within the span of branch 4 as seen in Figure~\ref{fig:mapn2}. Its positive growth rates are terminated by branch 4 where its phase reaches 
$180^{\circ}$, in a somewhat similar manner as occurs for modes that are a continuation of branch 2, as seen in the bottom three panels of Figure~\ref{fig:marginalmodes24}.
 
 Branches 1 and 4 cross
 at $h_{\rm out}=0.0654$ corresponding to $i=48.4^{\circ}$. 
 The second row of Figure~\ref{fig:marginalmodes24} shows the growth terminating
 by a crossing with branch 4 at $i=42.9^{\circ}$.
 Consequently, the boundary for the interaction between modes $M(2, h_{\rm out}, i)$  and branch 4 
 occurs at $h_{\rm out}$  values that are slightly larger  than where branches 1 and 4 cross, where the dashed short vertical line lies.

\begin{figure*}
\centering%
\includegraphics{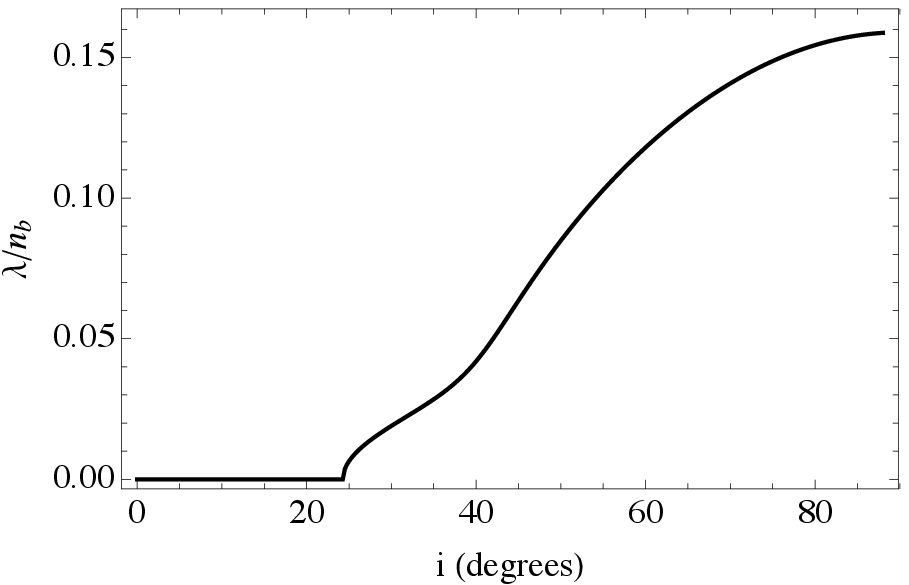}
\caption{Growth rate as a function of inclination angle for $h_{\rm out}=0.055$ for modes that extend from branch 2 of model A2.
 }
\label{fig:gr0p055}
\end{figure*}

Away from mode crossings, we find that  the  dominant (fastest growing) mode is the one that extends from 
the branch having the smallest  critical angle $i_{\rm crit}$. For example, for $h_{\rm out} \simeq 0.035$ to 0.07, modes
$M(2, h_{\rm out}, i)$ dominate over modes  $M(1, h_{\rm out}, i)$ for the same $h_{\rm out}$ and $i$, except possibly for the regions close to branch 4.
This can be seen for example 
in Figure~\ref{fig:gr0p055}. The plotted growth rates correspond to the same $h_{\rm out}=0.055$
value as for the upper left panel of  Figure~\ref{fig:modesn1}.  Positive growth rates for the mode  $M(1, 0.055,i)$
occur for $i \ga 46^{\circ}$. But at such angles the growth rates plotted in Figure~\ref{fig:gr0p055} for the modes $M(2, 0.055,i)$ are larger.

In Figure~\ref{fig:marginalA2B2} we see that the highest $h_{\rm out}$ for a marginal mode occurs for branch 1 that
reaches an inclination of $90^{\circ}$ for $h_{\rm out} \sim 0.15$. As a result, the range of unstable angles vanishes for such
disc aspect ratios and above.  Additional apsidal precession that is not due to the gravitational forces of the binary companion
can suppress KL oscillations if that additional apsidal precession rate exceeds that due to the binary 
\citep[e.g.,][]{Fabrycky07}.
As noted in \cite{Martin14}, the pressure-induced 
 apsidal precession rate is of order $\sim h_{\rm out}^2 n$.  For the pressure 
 apsidal precession rate  to exceed the binary precession rate $\sim q\, n_{\rm b}^2/n$ requires
 roughly $h_{\rm out} \ga \sqrt{q}\, n_{\rm b}/n$, where $q$ is the binary mass ratio $M_2/M_1$
 that is assumed to be in the range $q \la 1$, as is  similar to the findings of \cite{Zanazzi16}.  In the current case with $q=1$, we have that
$n_{\rm b}/n \sim 0.2$. So this relationship is roughly satisfied. 
As was also noted in \cite{Martin14}, another condition for KL oscillations is that the inverse radial sound
crossing time be shorter than the disc precession rate in order that the disc remains flat \citep{Larwood97}.
If this condition is not satisfied, the disc will severely warp and may break up.
This condition can crudely be written as $ n \, h_{\rm out}   \ga q n_{\rm b}^2/n$. For the models in this paper this condition crudely
requires that $h_{\rm out} \ga 0.05$. However, since we have dropped all factors of order unity, the 
flatness condition might be satisfied at  somewhat smaller values as well, as was found in SPH studies
\citep{Martin14, Fu15a}.
In any case, we then have crudely that KL oscillations can occur 
because there is a window of $h_{\rm out}$ values for which
\begin{equation}
q \left (\frac{n_{\rm b}}{n}\right)^{2} \la h_{\rm out} \la \sqrt{q} \, \frac{n_{\rm b}}{n}. \label{pwindow}
\end{equation}
This relationship can also be expressed as a constraint on binary mass ratios as
\begin{equation}
q \, h_{\rm out}\la q_{\rm crit} \la q \label{qwindow},
\end{equation}
where $q_{\rm crit} = (h_{\rm out} n/n_{\rm b})^2.$

For a disc that relies on only self-gravity to maintain flatness, the level of self-gravity required for  flatness
produces an apsidal precession rate that exceeds the precession rate due to the binary companion and therefore suppresses KL oscillations  \citep{Batygin11}.
As a result, there is no equivalent KL window analogous to Equation (\ref{pwindow}) for such a disc.

\section{Properties of KL Eccentric Modes} \label{sec:eigenprops}

As discussed in the Section \ref{sec:models}, the disc inner radius is likely to be about $\sim 10^{-4} a_{\rm b}$ 
 for wider binaries where the KL oscillation timescales are of order $10^{4}$ yr or larger.
We see from Figure~\ref{fig:marginalmodesA2B2} that the eccentricity at the inner boundary can be much larger
than the eccentricity in the outer parts of a disc.
We examine the behaviour of the eccentric modes at small radii and the sensitivity of the results to 
the inner radius.

For $r \ll a_{\rm b}$, the time dependence of $E$ can be ignored and the spatial 
dependence of $E$ can be determined.  Equation (\ref{E_eqiso}) reduces to
\begin{equation}
E''(r) + \frac{3 - p}{r} \, E'(r) + \frac{6 - p (1+ s) - (2 + s) s}{r^2} \, E(r)=0, \label{Esr}
\end{equation}
where $p$ and $s$ are the pressure and temperature exponents discussed in Section  \ref{sec:models}.
We apply boundary condition
\begin{eqnarray}
E'(r_{\rm in}) &= &0.
\end{eqnarray}
Equation  (\ref{Esr})  has power law solutions  in which the exponent is generally  a
complex number. For $\sigma$ real, terms of the form $r^{{\rm i} \,\sigma}$ can be expressed as $\exp{({\rm i}\, \sigma \log{r})}$.
For the range of parameters in this paper, the solution can then be expressed in the form
\begin{equation}
E(r)=\frac{\beta \cos{(k \log{(r/r_{\rm in})} - \phi)}}{(r/r_{\rm in})^z} \label{Eina}
\end{equation}
where $k$, $\phi$, and $z$ are real constants which depend on $p$ and $s$ and are determined analytically.
There is also a normalization constant $\beta$ that we adjust, as described below.
We find that
\begin{equation}
z=1-\frac{p}{2}. \label{z}
\end{equation}
For typical surface density profiles  $p<2$  and so $z$  is positive.  The eccentricity then varies at small $r$  as $e(r) \sim 1/r^z$, ignoring oscillatory terms,  
and is therefore divergent.

\begin{figure*}
\centering%
\includegraphics{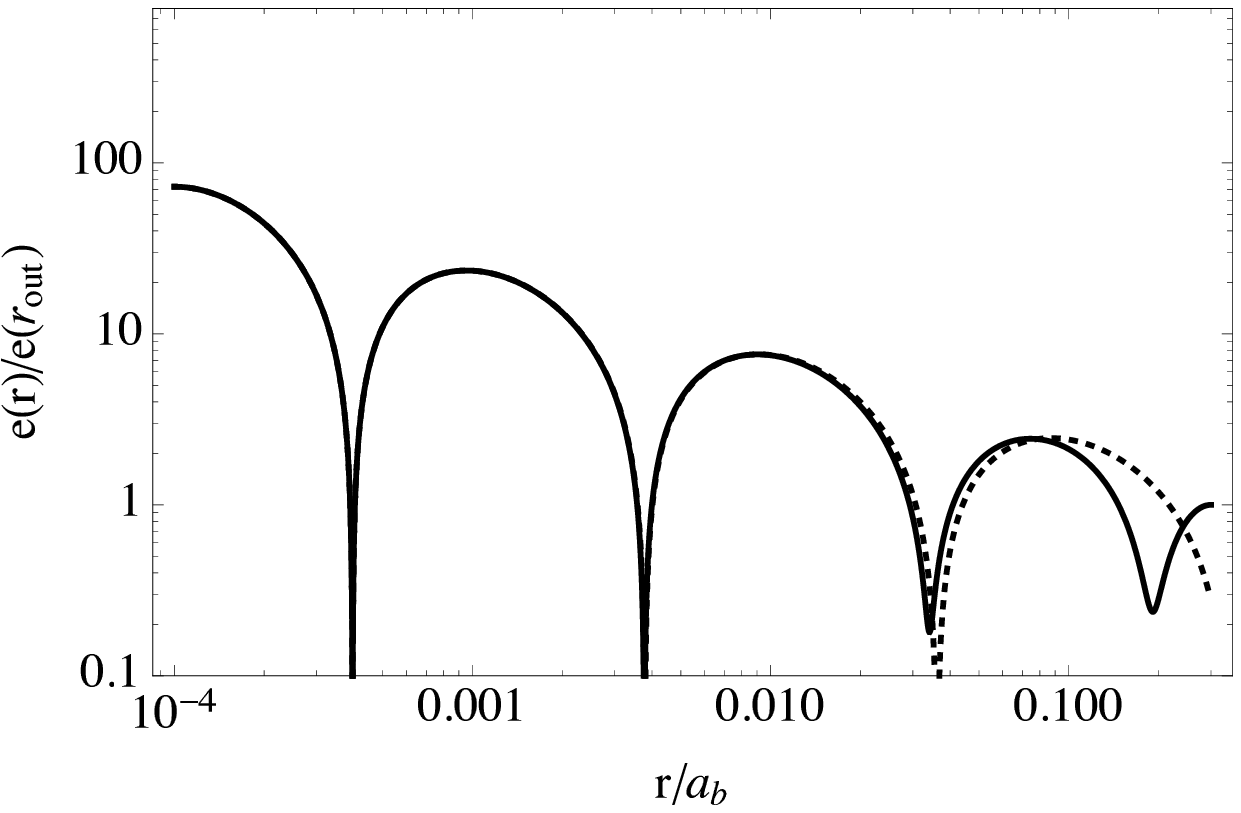}
\caption{Log-log plot of the eccentricity distribution for the model plotted in Figure~\ref{fig:gr0p055} with $i=30^{\circ}$.
The solid line is for the numerically determined result, while the dotted line is for the analytic approximation
given by Equation (\ref{Eina}) with parameters $k$, $\phi$, and $z$ determined analytically. Normalization constant $\beta$ is selected so that the two curves match at $r=r_{\rm in}=10^{-4} a_{\rm b}$.}
\label{fig:e30d}
\end{figure*}

For model A, the analytic solution of the form given in Equation (\ref{Eina}) has $k=\sqrt{31}/4$, $\phi=\arctan{(2/\sqrt{31}})$, and $z=1/2$. The only free parameter is the normalization constant $\beta$ that is chosen so that the analytic and numerically determined solutions match at $r=r_{\rm in}$.
Figure~\ref{fig:e30d} plots a typical case.  The analytic result is barely distinguishable from the numerical solution for 
$r \la 0.03 a_{\rm b}$. 

In  an adiabatic disc, there is a globally conserved quantity  called the angular momentum deficit, which expresses the difference between the angular momentum of elliptical and circular orbits with the same energy \citep{Goodchild06, Teyssandier16}.
In a locally isothermal disc, this quantity is not conserved
due to the exchange of heat with the background radiation.
However, there  is a  related globally conserved quantity that we refer to as the modified angular momentum
deficit (MAMD) that is given in equation~(C23) of \cite{Teyssandier16} as
\begin{equation}
{\cal A} =\int \frac{1}{2} |E|^2 \frac{\Sigma r^2 n}{T} \,  2 \pi r dr, \label{AMD}
\end{equation}
where the integral extends over the disc.
To measure the influence of the large amplitudes near the inner boundary, we determine the fractional contributions to ${\cal A}$ that arise in its vicinity. 

From Equation (\ref{Eina}), the MAMD per unit radius (the integrand in Equation (\ref{AMD})) can be shown to vary  as 
$\sim r^{s-1/2}$ (ignoring oscillatory terms). Consequently, the integrated contributions of the inner region
to the MAMD increase roughly as $\sim r^{s+1/2}$. This integral is then well behaved and nonsingular for small $r$ for positive values of $s$, even though
$|E(r)|=e(r)$ may be singular as $r=r_{\rm in}$ goes to zero. 

\begin{figure*}
\centering%
\includegraphics{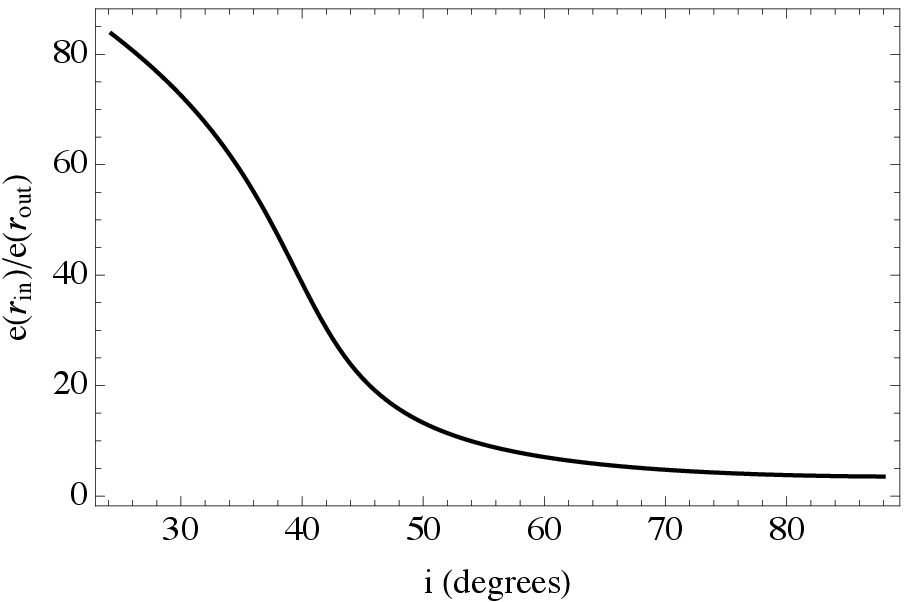}
\caption{Plot of the ratio of the eccentricity at  the inner boundary to that at the outer boundary for the model used in Figure~\ref{fig:gr0p055} as a function of inclination $i$.
}
\label{fig:ei0p055}
\end{figure*}

\begin{figure*}
\centering%
\includegraphics{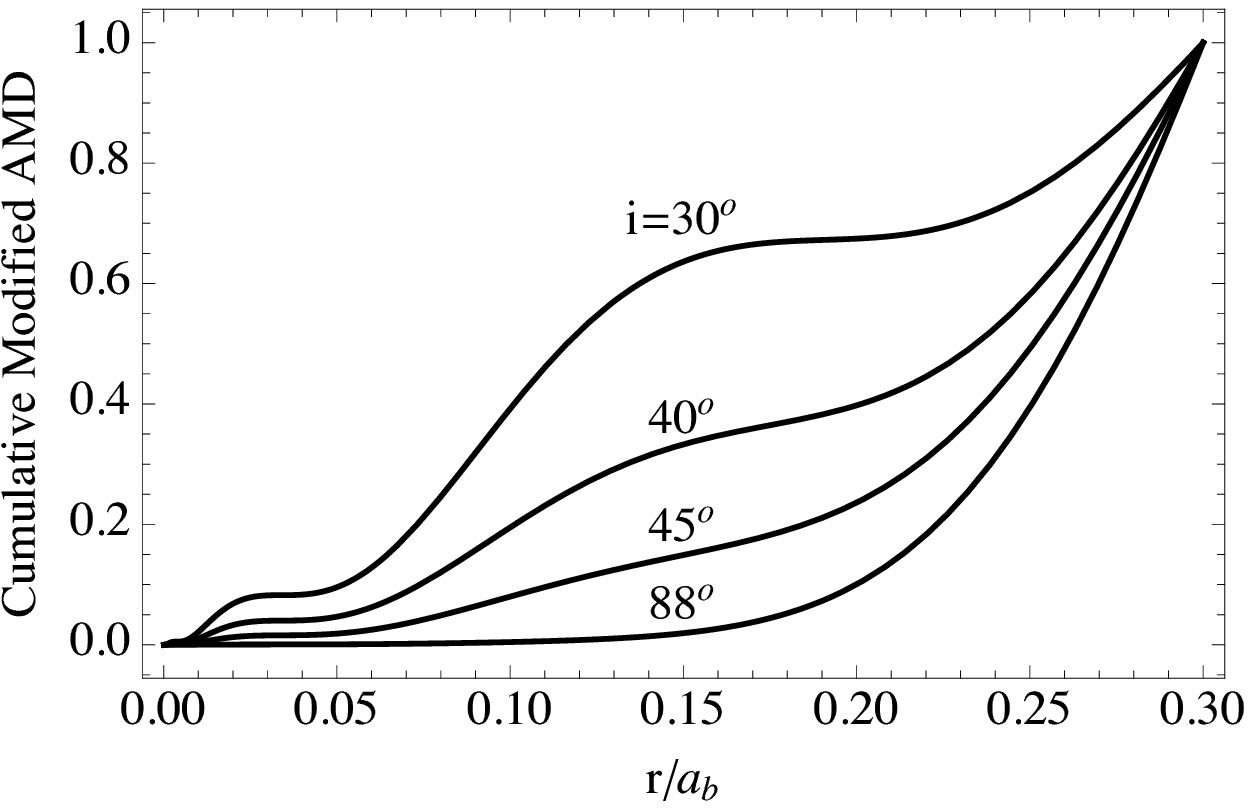}
\includegraphics{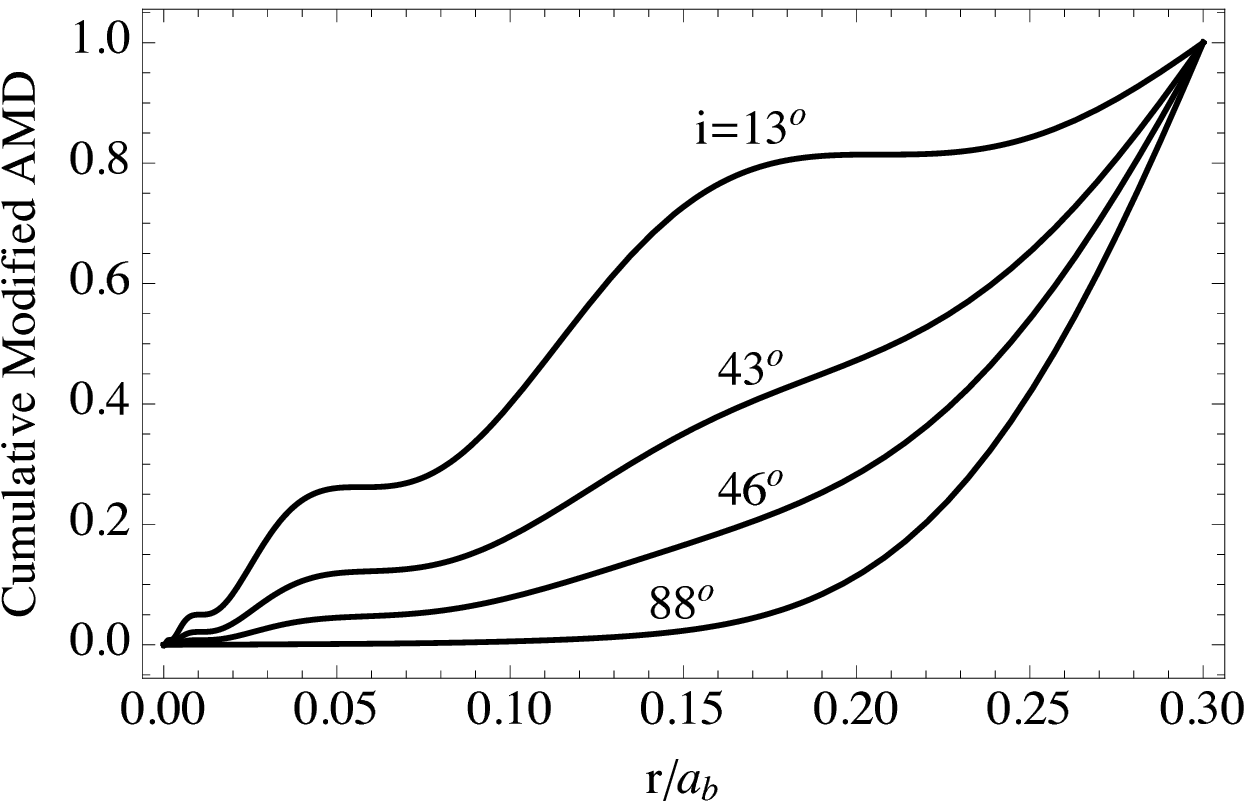}
\caption{Plots  of the cumulative radial distribution of modified angular momentum deficit for the model used  Figure~\ref{fig:gr0p055}. 
Top: Model A2 with $h_{\rm out}=0.055$. Bottom: Model B2 with $h_{\rm out}=0.059$.
}
\label{fig:AMD0p055}
\end{figure*}

Figure~\ref{fig:ei0p055} plots $e(r_{\rm in})/e(r_{\rm out})$ as a function of inclination $i$ for modes that extend from branch 2 of model A2 with $h_{\rm out}=0.055$.
At smaller inclinations, where the growth rate is positive but smaller (see Figure~\ref{fig:gr0p055}), the eccentricity at the inner boundary is substantially bigger than at the outer boundary. There is a rapid drop in this ratio near $i \simeq 42^{\circ}$.
In Figure~\ref{fig:AMD0p055} we plot the cumulative radial distribution starting at $r=r_{\rm in}$ of the MAMD. As expected from Figure~\ref{fig:ei0p055}, there is a rapid change in the distributions between $i=40^{\circ}$ and $45^{\circ}$. This distribution is normalized to unity at $r=r_{\rm out}$.
However, the cumulative MAMD distributions show only small contributions coming from the region near the inner boundary.  For both models the MAMD distributions are fairly similar at similar inclinations.
The MAMD near the inner boundary for $i=13^{\circ}$ in the bottom
panel reaches a few percent contribution out to $r \simeq 0.005 a_{\rm b}$ and the contributions are smaller at larger inclinations. 

\begin{figure*}
\centering%
\includegraphics{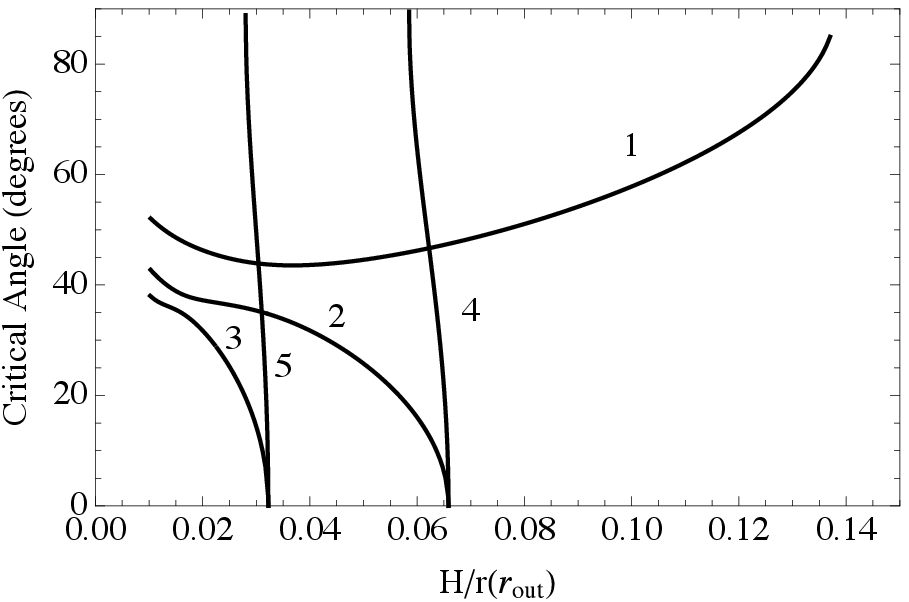}
\includegraphics{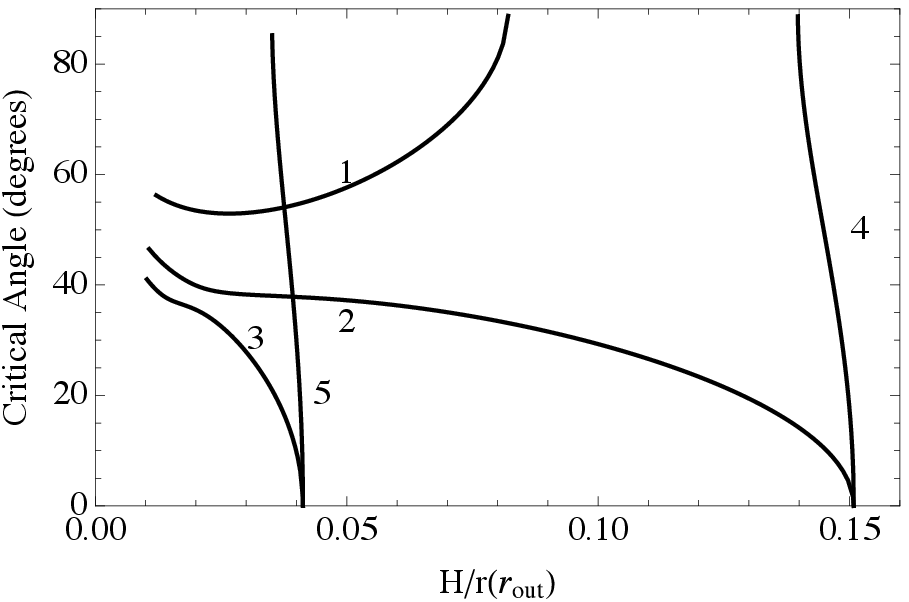}
\caption{Disc inclination $i$ as a function of disc aspect ratio $H/r(r_{\rm out})$ for marginally stable modes
of disc model A1 (top panel) and model B1 (bottom panel). The branches are labeled by integers. }
\label{fig:marginalA1B1}
\end{figure*}

We consider the sensitivity of the results to the inner boundary location by comparing model A1 to A2 and models B1 to B2.
The models being compared have inner radii that differ by a factor of 100. Figure~\ref{fig:marginalA1B1} plots the branches
of marginal modes for models A1 and B1. In comparing this figure with Figure~\ref{fig:marginalA2B2} we see that 
there are only small differences between models A1 and A2.

\begin{figure*}
\centering%
\includegraphics[width=10cm]{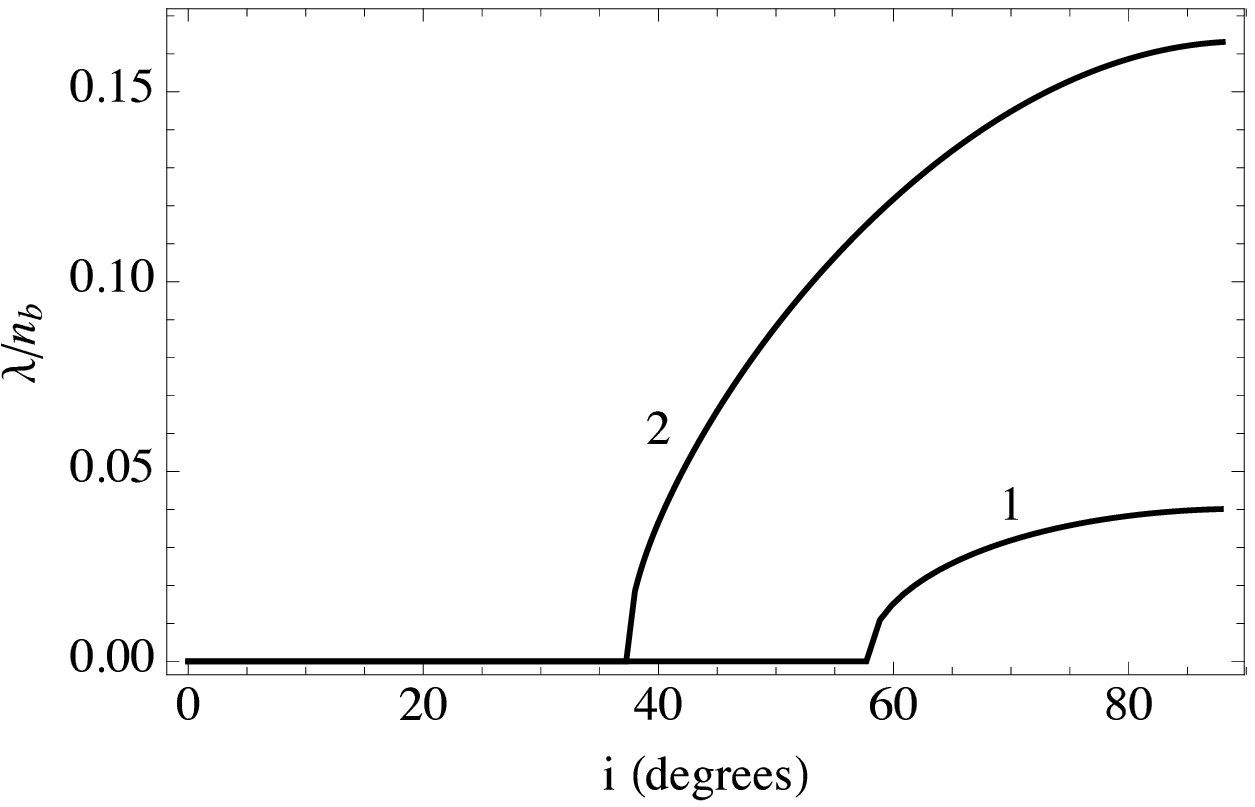}
\caption{Plot of the growth rates as a function of inclination for $h_{\rm out} =0.05$  for modes that lie on extensions of branches 1 and 2 of model B1.
}
\label{fig:grB1}
\end{figure*}

On the other hand, there are major differences between models B1 and B2. 
Model B1 has some unusual properties.
Branch 1  (the branch that does not reach to $i=0$) ranges over lower
$h_{\rm out}$ values in model B1 than in model B2.
A major difference is that modes that lie on the extension of branch 1 in model B1
never dominate. The reason is that the span of $h_{\rm out}$ values covered by branch 1 is also covered by other branches with lower 
critical angles for marginal stability.  For example, at $h_{\rm out}=0.05$ branches 1 and 2 are marginally stable.
As shown in Figure~\ref{fig:grB1}, the modes that extend from branch 2 dominate at all inclinations in this case. The termination of positive growth with increasing $h_{\rm out}$  in model B1
is then different from the other models we have presented. Rather than growth being terminated at higher $h_{\rm out}$ by branch 1 at 
$i_{\rm crit}=90^{\circ}$,
as in the other models, growth in model B1 appears to be  terminated by branch 2 at $i_{\rm crit}=0^{\circ}$ that occurs for $h_{\rm out} \simeq 0.15$.

For a locally isothermal disc with a very small inner radius, we estimate the  
scaling of the eccentricity required for nonlinearity to set in.
A relevant measure of nonlinearity is $|r dE/dr-E|$ \citep{Ogilvie01}.  
Using Equation (\ref{z}), we have that this measure of nonlinearity in the inner disc scales like $e(r) \sim e(r_{\rm out}) r^{p/2-1}$.  So for fixed $i$, the radius within which the mode becomes nonlinear scales as $r_{\rm NL} \sim e(r_{\rm out})^{1/(1-p/2)}$.  The fraction of MAMD that is subject to nonlinear modification scales as  $\sim r_{\rm NL}^{s+1/2} \sim e(r_{\rm out})^{(2s+1)/(2-p)}$.  
For models A and B this power is 2.5 and 2 respectively. Nonlinearity of the eccentric mode can lead to enhanced dissipation.  When the eccentricity gradient is sufficiently large that neighbouring orbits approach mutual intersection (Ogilvie 2001), the local rate of viscous damping of the mode increases nonlinearly.  At slightly lower amplitudes, inertial waves in the disc become strongly destablilized by the eccentric motion \citep{Papaloizou05a, Papaloizou05b, Barker14}, and this is also expected to damp the eccentricity in a way that depends nonlinearly on its amplitude. The occurrence of either of these mechanisms in the inner disc provides a possible way to saturate the growth of a global mode through nonlinear damping, regardless of the (very small) value of the inner radius.

\section{Summary}
We have investigated the linear stability of tilted discs in  binary star systems
to Kozai--Lidov (KL) oscillations by extending the model of \cite{Teyssandier16} that includes 3D effects in a 1D calculation.
The results support the existence of disc KL oscillations found in SPH simulations \citep{Martin14, Fu15a, Fu15b}.
We find that sufficiently tilted discs can undergo KL oscillations in binaries with order unity mass ratios provided that roughly  $(n_{\rm b}/n)^2 \la h_{\rm out} \la n_{\rm b}/n$, for binary orbital frequency $n_{\rm b}$, disc aspect ratio $h_{\rm out}$, and  orbital frequency $n$
 at the disc outer edge.
In agreement with \cite{Zanazzi16}, we find KL disc instability is not possible for binaries with 
 small mass perturbers in which the binary mass ratio $M_2/M_1 \la (h_{\rm out} \, n/n_{\rm b})^2$, but that
 instability is possible  in equal mass binaries for disc inclinations that lie
  below the  critical value required for test particles of $39.2^{\circ}$.

As the disc evolves, its eccentricity follows an eigenmode (see Figure~\ref{fig:timeint}). 
The states of marginal stability lie on mode branches 
in a $h_{\rm out}-i_{\rm crit}$ diagram, as shown in Figure~\ref{fig:marginalA2B2}. Generally, discs are unstable if their tilts lie above the critical inclination for any marginal stability branch. There are cases where this is not true, as seen in the lower panel of Figure~\ref{fig:marginalA1B1}.  Modes can be unstable in discs that
lie very close to the disc midplane, unlike the case of test particles. However, the growth rates and range of unstable angles is small for nearly coplanar discs
(see lower panels of Figure~\ref{fig:marginalmodes24}). The coplanar  ($i=0$) apsidal-nodal disc resonances
delineate the marginally stable branches. The disc aspect ratios $h_{\rm out}$ for these resonances are a function
of the disc temperature and density gradients (see Figures \ref{fig:tracks} and \ref{fig:trackp}).
In general,
more than one unstable mode can be present in a disc. The number of these unstable modes increases in cooler discs. Mode crossings limit the range of unstable tilt angles for modes that lie on extensions of branches whose critical inclinations $\la 12^{\circ}$,
as seen in Figure~\ref{fig:marginalmodes24}.
 The dominant mode is the one that lies on an extension of a marginally stable branch having with the smallest critical angle (e.g., Figure~\ref{fig:grB1}).
 
 Discs at smaller tilt angles tend to have larger eccentricities near the disc inner edge relative to the disc outer edge (Figure~\ref{fig:ei0p055}). The eccentricities are formally divergent as the inner radius goes to zero.  However, their  
 effect on the
modified angular momentum deficit near the inner edge is small (Figure~\ref{fig:AMD0p055}).
At the end of Section 8, we describe the scaling of eccentricity for nonlinearity to set in.

We found some sensitivity of the results to the location of the inner boundary in models B1 and B2.   The marginal stability curves, especially for  branch 1 for model B1 in Figure~\ref{fig:marginalA1B1}, are shifted relative to model B2  
in Figure~\ref{fig:marginalA2B2}.
 
We investigated some of the interactions between modes at mode crossings (see Figure~\ref{fig:mapn2}). But more remains
to be explored, particularly involving oscillatory unstable modes.
We have idealized the disc as flat, but bending
modes may be present that could provide additional mechanisms for instability.

\section*{Acknowledgements}
We thank Rebecca Martin for useful discussions
and informing us about the preprint by \cite{Zanazzi16}.
SHL acknowledges support from NASA grant NNX11AK61G.





\begin{thebibliography}{99}
\bibitem[Barker \& Ogilvie (2014)]{Barker14}
Barker, A. J., \& Ogilvie, G. I. 2014, \mnras, 445, 2637

\bibitem[Bate et al.\ (2000)]{Bate00}
Bate, M. R., Bonnell, I. A., Clarke, C. J., et al. 2000, MNRAS, 317, 773
\bibitem[Bate et al.\ (2010)]{Bate10}
Bate, M. R., Lodato, G., \& Pringle, J. E. 2010, MNRAS, 401, 1505
\bibitem[Batygin et al.\ (2011)]{Batygin11} Batygin, K., Morbidelli, A., \& Tsiganis, K.\ 2011, AA, 533, A7
\bibitem[Chen et al.\ (2011)]{Chen11}
Chen, X., Sesana, A., Madau, P., \& Liu, F. K. 2011, ApJ, 729, 13
\bibitem[Chiang \& Goldreich (1997)]{Chiang97}
Chiang, E.I. \& Goldreich, P.  1997, ApJ, 490, 368
\bibitem[Fabrycky \& Tremaine (2007)]{Fabrycky07}
Fabrycky, D., \& Tremaine, S. 2007, ApJ, 669, 1298
\bibitem[Ford et al.\ (2000)]{Ford00}
Ford, E. B., Kozinsky, B., \& Rasio, F. A. 2000, ApJ, 535, 385
\bibitem[Fu et al.\ (2015a)]{Fu15a} Fu, W., Lubow, S.~H., \& Martin, R.~G.\ 2015a, ApJ, 807, 75 
\bibitem[Fu et al.\ (2015b)]{Fu15b} Fu, W., Lubow, S.~H., \& Martin, R.~G.\ 2015b, ApJ, 813, 105 
\bibitem[Fu et al.\ (2016)]{Fu16} Fu, W., Lubow, S.~H., \& Martin, R.~G.\ 2016, arXiv:1612.07673
\bibitem[Goodchild \& Ogilvie (2006)]{Goodchild06} Goodchild, S., \& Ogilvie, G. 2006, MNRAS, 381 
\bibitem[Hale (1994)]{Hale94} Hale, A. 1994, AJ, 107, 306
\bibitem[Jensen \& Akeson (2014)]{Jensen14} Jensen, E. L. N. \& Akeson, R. 2014, Nature, 511, 567
\bibitem[King et al.\ (2013)]{King13} King, A. R., Livio, M., Lubow, S. H., \& Pringle, J. E. 2013, MNRAS, 431, 2655
\bibitem[Kiseleva et al.\ (1998)]{Kiseleva98} Kiseleva, L. G., Eggleton, P. P., \& Mikkola, S. 1998, MNRAS, 300, 292
\bibitem[Kozai (1962)]{Kozai62} Kozai, Y. 1962, AJ, 67, 591
\bibitem[Kushnir et al.\ (2013)]{Kushnir13} Kushnir, D., Katz, B., Dong, S., Livne, E., \& Fern\'{a}ndez, R. 2013, ApJ, 778, L37
\bibitem[Larwood \& Papaloizou (1997)]{Larwood97} Larwood, J. D., \& Papaloizou, J. C. B. 1997, MNRAS, 285, 288
\bibitem[Li et al.\ (2014)]{Li14}
Li, G., Naoz, S., Kocsis, B., \& Loeb, A. 2014, ApJ, 785, 116
\bibitem[Lidov (1962)]{Lidov62}
Lidov, M. L. 1962, P\&SS, 9, 719
\bibitem[Lin, Bodenheimer, \& Richardson (1996)]{Lin96}  Lin, D.N.C., Bodenheimer, P.,
\& Richardson, D.C. 1996, Nature, 380, 606
\bibitem[Lithwick \& Naoz (2011)]{Lithwick11}
Lithwick, Y., \& Naoz, S. 2011, ApJ, 742, 94
\bibitem[Liu et al.\ (2015)]{Liu15}
Liu, B., Mu\~{n}oz, D., \& Lai, D. 2015, MNRAS, 447, 747
\bibitem[Lubow  (1991)]{Lubow91}
Lubow, S. H. 1991, ApJ, 381, 259

\bibitem[Lubow \& Ogilvie (2000)]{Lubow00}
Lubow, S. H., \& Ogilvie, G. I. 2000, ApJ, 538, 326

\bibitem[Lubow \& Ogilvie (2001)]{Lubow01}
Lubow, S. H., \& Ogilvie, G. I. 2001, ApJ, 560, 997

\bibitem[Lubow (2010)]{Lubow10}
Lubow, S. H. 2010, MNRAS, 406, 2777

\bibitem[Lubow, Martin \& Nixon (2015)]{Lubow15}
Lubow, S. H., Martin, R. G., \& Nixon, C. J. 2015, ApJ, 800, 96

\bibitem[Lynden-Bell \& Pringle (1974)]{Lynden-Bell74}
Lynden-Bell, D. \& Pringle, J. E., MNRAS, 168, 603

\bibitem[Martin et al.\ (2011)]{Martin11}
Martin R. G.,  Pringle J. E., Tout C. A., \& Lubow S. H. 2011, MNRAS, 416, 2827

\bibitem[Martin et al.\ (2014)]{Martin14}
Martin, R. G., Nixon, C., Lubow, S. H., et al. 2014b, ApJ, 792, L33

\bibitem[{{Miranda} \& {Lai}(2015)}]{Miranda2015}
{Miranda}, R. \& {Lai}, D. 2015, MNRAS, 452, 2396

\bibitem[Naoz et al. (2013)]{Naoz13}
Naoz, S., Kocsis, B., Leob A., \& Yunes, N. 2013b, ApJ,773, 187

\bibitem[Nesvorn\'{y} et al.\ (2003)]{Nesvorny03}
Nesvorn\'{y}, D., Alvarellos, J. L. A., Dones, L., \& Levison, H. F. 2003, AJ, 126, 398

\bibitem[{{Nixon} \& {Lubow}(2015)}]{Nixon15}
{Nixon}, C. \& {Lubow}, S.~H. 2015, MNRAS, 448, 3472

\bibitem[Ogilvie(1999)]{Ogilvie99} Ogilvie, G.~I.\ 1999, MNRAS, 304, 557 

\bibitem[Ogilvie (2001)]{Ogilvie01}
Ogilvie, G. I. 2001, MNRAS, 325, 231

\bibitem[Ogilvie \& Dubus (2001)]{Ogilvie01a}
Ogilvie, G. I. \& Dubus, G. 2001, \mnras, 320, 48

\bibitem[Ogilvie(2006)]{Ogilvie06} Ogilvie, G.~I.\ 2006, MNRAS, 365, 977 

\bibitem[Ogilvie(2008)]{Ogilvie08} Ogilvie, G.~I.\ 2008, MNRAS, 388, 1372 

\bibitem[Paczy\'{n}ski (1977)]{Paczynski77}
Paczy\'{n}ski, B. 1977, ApJ, 216, 822


\bibitem[Papaloizou \& Terquem (1995)]{Papaloizou95}
Papaloizou, J. C. B., \& Terquem, C. 1995, MNRAS, 274, 987

\bibitem[Papaloizou (2005a)]{Papaloizou05a}
Papaloizou, J. C. B. 2005a, A\&A, 432, 743

\bibitem[Papaloizou (2005b)]{Papaloizou05b}
Papaloizou, J. C. B. 2005b, A\&A, 432, 757

\bibitem[Pringle (1996)]{Pringle96}
Pringle, J. E. 1996, \mnras, 281, 357

\bibitem[Stapelfeldt et al.\ (1998)]{Stapelfeldt98}
Stapelfeldt, K. R., Krist, J. E., M\'{e}nard, F., et al. 1998, ApJ, 502, L65


\bibitem[Teyssandier et al.\ (2013)]{Teyssandier13}
Teyssandier, J., Naoz, S., Lizarraga, I., \& Rasio, F. A. 2013, ApJ, 779, 166

\bibitem[Teyssandier \& Ogilvie (2016)]{Teyssandier16} Teyssandier, J., \& Ogilvie, G.~I.\ 2016, MNRAS, 458, 3221

\bibitem[Tremaine \& Yavetz (2014)]{Tremaine14} Tremaine, S., \& Yavetz, T.~D.\ 2014, American Journal of Physics, 82, 769 

\bibitem[Williams et al.\ (2014)]{Williams14}
Williams, J. P., Mann, R. K., Di Francesco, J., et al. 2014, ApJ, 796, 120

\bibitem[Willams \& Cieza (2011)]{Williams11}
Willams,  J. P. \& Cieza, L. A. 2011, ARAA, 49, 67

\bibitem[Zanazzi \& Lai (2016)]{Zanazzi16} Zanazzi,J. J. \& Lai, D. 2016, arXiv:1612.05598v1

\end{thebibliography}







\bsp	
\label{lastpage}
\end{document}